\documentclass{article}

\usepackage{neurips_2019}

\usepackage[utf8]{inputenc} 
\usepackage[T1]{fontenc}    
\usepackage{hyperref}       
\usepackage{url}            
\usepackage{booktabs}       
\usepackage{amsfonts}       
\usepackage{amsmath}
\usepackage{amssymb}
\usepackage{nicefrac}       
\usepackage{microtype}      

\usepackage[pdftex]{graphicx}
\usepackage[all]{hypcap}

\usepackage{xcolor}
\usepackage{soul}

\usepackage{bm}
\renewcommand{\v}{\bm}

\usepackage{lineno}

\title{Artificial neural networks for neuroscientists: \\A primer}

\author{\textbf{Guangyu Robert Yang$ ^{1,*} $, Xiao-Jing Wang$ ^{2,*} $} \\
$^{1}$ Center for Theoretical Neuroscience, Columbia University\\
$^{2}$ Center for Neural Science, New York University  \\ 
$^{*}$ Correspondence robert.yang@columbia.edu, xjwang@nyu.edu}

\begin{document}

\maketitle

\begin{abstract}
    Artificial neural networks (ANNs) are essential tools in machine learning that have drawn increasing attention  in neuroscience. Besides offering powerful techniques for data analysis, ANNs provide a new approach for neuroscientists to build models for complex behaviors, heterogeneous neural activity and circuit connectivity, as well as to explore optimization in neural systems, in ways that traditional models are not designed for. In this pedagogical Primer, we introduce ANNs and demonstrate how they have been fruitfully deployed to study neuroscientific questions. We first discuss basic concepts and methods of ANNs. Then, with a focus on bringing this mathematical framework closer to neurobiology, we detail how to customize the analysis, structure, and learning of ANNs to better address a wide range of challenges in brain research.  To help the readers garner hands-on experience, this Primer is accompanied with tutorial-style code in PyTorch and Jupyter Notebook, covering major topics.
\end{abstract}

\section{Artificial neural networks in neuroscience}
Learning with artificial neural networks (ANNs), or deep learning, have emerged as a dominant framework in machine learning (ML) nowadays \citep{lecun2015deep}, leading to breakthroughs across a wide range of applications, including computer vision \citep{krizhevsky2012imagenet}, natural language processing \citep{devlin2018bert}, and strategic games \citep{silver2017mastering}. Some key ideas in this field can be traced to brain research: supervised learning rules have their roots in the theory of training perceptrons which in turn was inspired by the brain \citep{rosenblatt1962principles}; the hierarchical architecture \citep{fukushima1982neocognitron} and convolutional principle \citep{lecun95} were closely linked to our knowledge about the primate visual system \citep{hubel62,felleman1991distributed}. Today, there is a continued exchange of ideas from neuroscience to the field of artificial intelligence \citep{hassabis2017neuroscience}. 
 
At the same time, machine learning offers new and powerful tools for systems neuroscience. One utility of the deep learning framework is to analyze neuroscientific data (Figure 1). Indeed, the advances in computer vision, especially convolutional neural networks, have revolutionized image and video data processing. For instance, uncontrolled behaviors over time, such as micro-movements of animals in a laboratory experiment, can now be tracked and quantified efficiently with the help of deep neural networks \citep{mathis2018deeplabcut}. Innovative neurotechnologies are producing a deluge of big data from brain connectomics, transcriptome and neurophysiology, the analyses of which can benefit from machine learning.  Examples include image segmentation to achieve detailed, $\mu$m scale, reconstruction of connectivity in a neural microcircuit  \citep{januszewski2018high,helmstaedter2013connectomic}, and estimation of neural firing rate from spiking data \citep{pandarinath2018inferring}.
 
This primer will not be focused on data analysis; instead, our primary aim is to present basic concepts and methods for the development of ANN models of biological neural circuits in the field of computational neuroscience. It is noteworthy that ANNs should not be confused with neural network models in general. Mathematical models are all ``artificial" inasmuch as they are not biological. We denote by ANNs specifically models that are in part inspired by neuroscience yet for which biologically justification is not the primary concern, in contrast to other types of models that strive to be built on quantitative data from the two pillars of neuroscience: neuroanatomy and neurophysiology. The use of ANNs in neuroscience \citep{zipser1988back} and cognitive science \citep{cohen1990control} dates back to the early days of ANNs \citep{rumelhart1986learning}. In recent years, ANNs are becoming increasingly common model systems in neuroscience \citep{yamins2016using,kriegeskorte2015deep,sussillo2014neural,barak2017recurrent}. There are three reasons for which ANNs or deep learning models have already been, and will likely continue to be, particularly useful for  neuroscientists.

\begin{figure}
  \centering
  \includegraphics[width=1.0\linewidth]{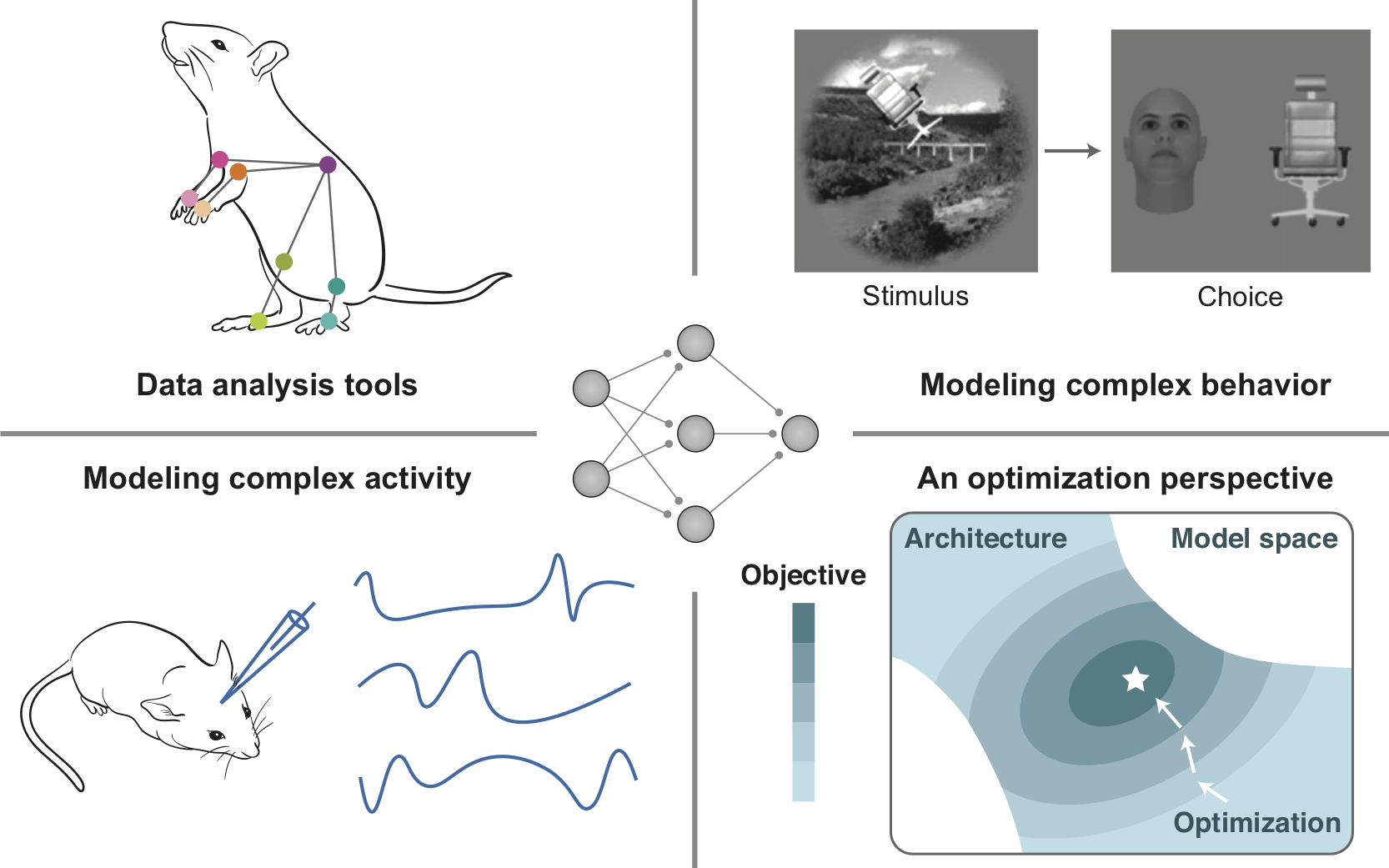}
  \caption{\textbf{Reasons for using ANNs for neuroscience research.} (Top left) Neural/Behavioral data analysis. ANNs can serve as image processing tools for efficient pose estimation (color dots). Figure inspired from \cite{nath2019using}. (Top right) Modeling complex behaviors. ANNs can perform object discrimination tasks involving challenging naturalistic visual objects. Figure adapted from \cite{kar2019evidence}. (Bottom left) Illustrating that ANNs can be used to model complex neural activity/connectivity patterns (blue lines). (Bottom right) Understanding neural circuits from an optimization perspective. In this view, functional neural networks (star symbol) are results of the optimization (arrows) of an objective function in an abstract space of a model constrained by the neural network architecture (colored space).}
  \label{fig:intro}
\end{figure}

First, fresh modeling approaches are needed to meet new challenges in brain research. Over the past decades, computational neuroscience has made great strides and become an integrated part of systems neuroscience \citep{abbott2008theoretical}. Much insights have been gained through integration of experiments and theory, including the idea of excitation and inhibition balance \citep{van1996chaos,shu2003turning} and normalization \citep{carandini2012normalization}. Progress was also made in developing models of basic cognitive functions such as simple decision-making \citep{gold2007neural,wang2008decision}. However, real-life problems can be incredibly complex, the underlying brain systems are often difficult to capture with ``hand-constructed'' computational models. For example, object classification in the brain is carried out through many layers of complex linear-nonlinear processing. Building functional models of the visual systems that achieve behavioral performance close to humans' remained a formidable challenge not only for neuroscientists, but also for computer vision researchers. By directly training neural network models on complex tasks and behaviors, deep learning provides a way to efficiently generate candidate models for brain functions that otherwise could be near impossible to model (Figure 1). By learning to perform a variety of complex behaviors of animals, ANNs could serve as potential model systems for biological neural networks, complementing nonhuman animal models for understanding the human brain.
 
A second reason for advocating deep networks in systems neuroscience is the acknowledgment that relatively simple models often do not account for a wide diversity of activity patterns in heterogeneous neural populations (Figure 1). One can rightly argue that this is a virtue rather than defect because simplicity and generality are hallmarks of good theories. However, complex neural signals also tell us that existing models may be insufficient to elucidate mysteries of the brain. This is perhaps especially true in the case of the prefrontal cortex. Neurons in prefrontal cortex often show complex mixed selectivity to various task variables \citep{rigotti2010internal,rigotti2013importance}. Such complex patterns are often not straightforward to interpret and understand using hand-built models that by design strive for simplicity. ANNs are promising to capture the complex nature of neural activity.

Thirdly, besides providing mechanistic models of biological systems, machine learning can be used to probe the ``why'' question in neuroscience \citep{barlow1961possible}. Brains are biological machines evolved under pressure to compute robustly and efficiently. Even when we understand how a system works, we may still ask \textit{why} it works that way. Similarly to biological systems evolving to survive, ANNs are trained to optimize objective functions given various architectural constraints (the number of neurons, economy of circuit wiring, etc.) (Figure 1). By identifying the particular objective and set of constraints that lead to brain-resembling ANNs, we could potentially gain insights into the evolutionary pressure faced by biological systems \citep{richards2019deep}.

In this pedagogical primer, we will discuss how ANNs can benefit neuroscientists in the three ways described above. In section 2, we will first introduce the key ingredients common in any study of ANNs. In section 3, we will describe two major applications of ANNs as neuroscientific models: convolutional networks as models for sensory, especially visual, systems, and recurrent neural networks as models for cognitive and motor systems. In the following sections 4 and 5, we will overview how to customize the analysis and architectural design of ANNs to better address a wide range of neuroscience questions. To help the readers gain hands-on experience, we accompany this primer with tutorial-style code in PyTorch and Jupyter Notebook (\url{https://github.com/gyyang/nn-brain}), covering all major topics.

\section{Basic ingredients and variations in artificial neural networks}
In this section, we will introduce basic concepts in ANNs and their common variations. Readers can skip this section if they are familiar with ANNs and deep learning. For a more thorough introduction, readers can refer to \cite{goodfellow2016deep}.

\subsection{Basic ingredient: learning problem, architecture, and algorithm}
A typical study using deep networks consists of three basic ingredients: learning problem, network architecture, and training algorithm. Weights of connections between units or neurons in a neural network are constrained by the network architecture, but their specific values are randomly assigned at initialization. These weights constitute a large number of parameters, collected denoted by $\v{\theta}$ which also includes other model parameters (see below), to be trained using an algorithm. The training algorithm specifies how connection weights change to better solve a learning problem, such as to fit a dataset or perform a task. We will go over a simple example, where a multi-layer-perceptron (MLP) is trained to perform a simple digit-classification task using supervised learning.

\paragraph{Learning problem} In supervised learning, a system learns to fit a dataset containing a set of inputs $\{\v{x}^{(i)}\}, i=1,\cdots, N$. Each input $\v{x}^{(i)}$ is paired with a target output $\v{y}_{\rm target}^{(i)}$. Symbols in bold represent vectors (column vectors by default). The goal is to learn parameters $\v{\theta}$ of a neural network function $F(\cdot, \v{\theta})$ that predicts the target outputs given inputs, $\v{y}^{(i)} = F(\v{x}^{(i)}, \v{\theta})\approx \v{y}_{\rm target}^{(i)}$. In the simple digit-classification task MNIST \citep{lecun1998gradient}, each input is an image containing a single digit, while the target output is a probability distribution over all classes (0, 1, ..., 9) given by a 10-dimensional vector or simply an integer corresponding to the class of that object.

More precisely, the system is trained to optimize the value of an objective function, or commonly, minimize the value of a loss function $L=\frac{1}{N}\sum_i L(\v{y}^{(i)}, \v{y}_{\rm target}^{(i)})$, where $L(\v{y}^{(i)}, \v{y}_{\rm target}^{(i)})$ quantifies the difference between the target output $\v{y}_{\rm target}^{(i)}$ and the actual output $\v{y}^{(i)}$.

\paragraph{Network architecture}
ANNs are incredibly versatile, including a wide range of architectures. Of all architectures, the most fundamental one is a Multi-Layer Perceptron (MLP) \citep{rosenblatt1958perceptron,rosenblatt1962principles} (Figure 2A). A MLP consists of multiple layers of neurons, where neurons in the $l$-th layer only receive inputs from the $(l-1)$-th layer, and only project to the $(l+1)$-th layer.
\begin{align}
    \v{r}^{(1)} & = \v{x}, \label{eq:forwardpass1}\\
    \v{r}^{(l)} & = f(\v{W}^{(l)} \v{r}^{(l-1)} + \v{b}^{(l)}), \ 1 < l < N, \\
    \v{y} & = \v{W}^{(N)} \v{r}^{(N-1)} + \v{b}^{(N)}. \label{eq:forwardpass3}
\end{align} 
Here $\v{x}$ is an external input, $\v{r}^{(l)}$ denotes the neural activity of neurons in the $l$-th layer, and $\v{W}^{(l)}$ is the connection matrix from the $(l-1)$-th to the $l$-th layer. $f(\cdot)$ is a (usually nonlinear) activation function of the model neurons. The output of the network is read out through connections $\v{W}^{(N)}$. Parameters $\v{b}^{(l)}$ and $\v{b}^{(N)}$ are biases for model neurons and output units respectively. If the network is trained to classify, then the output is often normalized such that $\sum_j y_j = 1$, where $y_j$ represents the predicted probability of class $j$.

When there are enough neurons per layer, MLPs can in theory approximate arbitrary functions \citep{hornik1989multilayer}. However, in practice, the network size is limited, and good solutions may not be found through training even when they exist. MLPs are often used in combination with, or as parts of, more modern neural network architectures.

\begin{figure}
  \centering
  \includegraphics[width=1.0\linewidth]{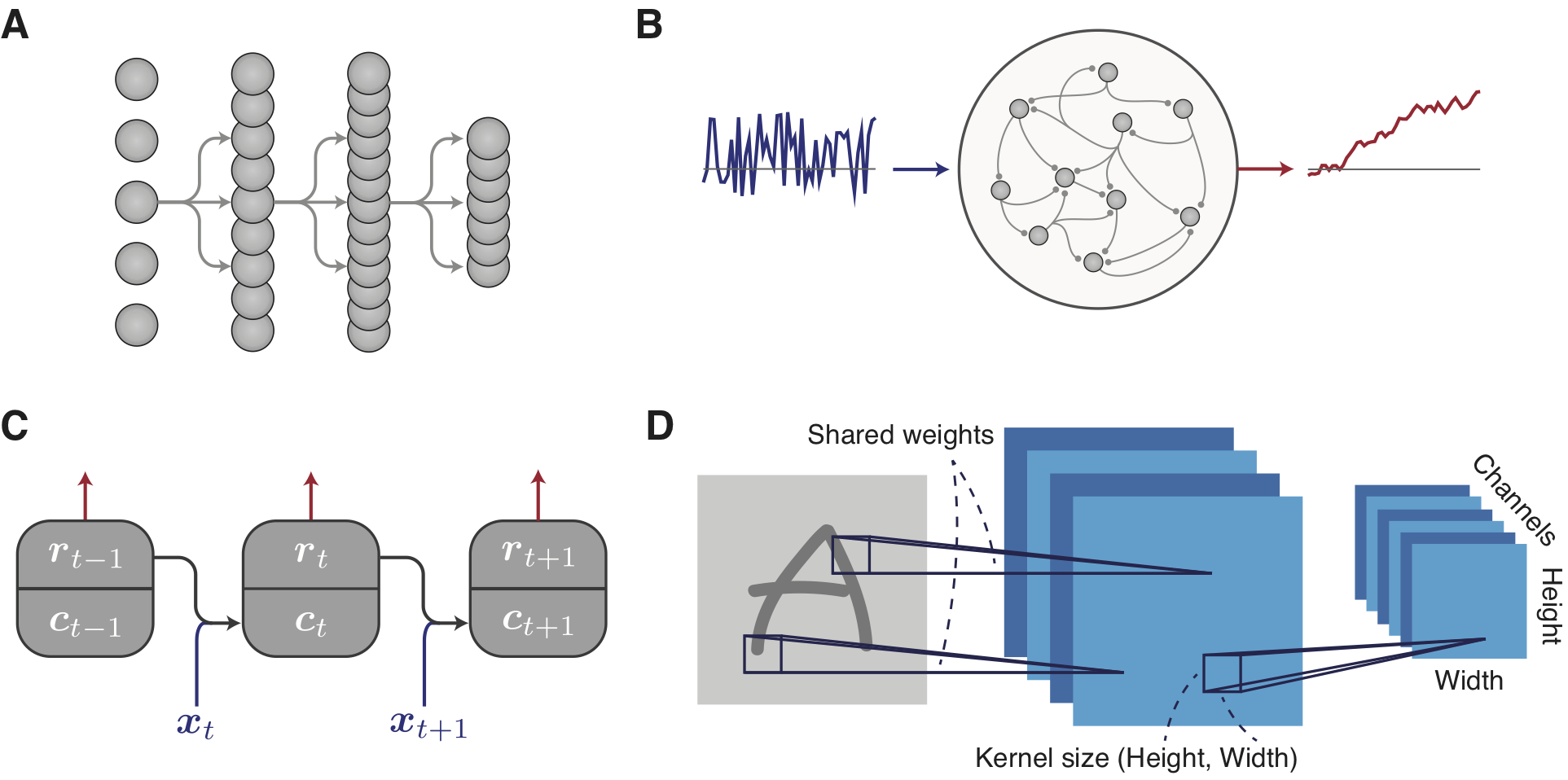}
  \caption{\textbf{Schematics of common neural network architectures.} (A) A multi-layer perceptron (MLP). (B) A recurrent neural network (middle) receives a stream of inputs (left). After training, an output unit (right) should produce a desired output. Figure inspired from \cite{mante2013context}. (C) A recurrent neural network is unrolled in time as a feedforward system with each layer corresponding to the network state at one time step. $\v{c}_t$ and $\v{r}_t$ describe the network state and output activity at time $t$ respectively. $\v{c}_t$ is a function of $\v{r}_{t-1}$ and the input $\v{x}_t$.
  (D) A convolutional neural network for processing images. Each layer contains a number of channels (4 in layer 1, 6 in layer 2). A channel (represented by a square) consists of spatially organized neurons, each receiving connections from neurons with similar spatial preferences. The spatial extent of these connections is described by the kernel size. Figure inspired from \cite{lecun1998gradient}.}
  \label{fig:architectures}
\end{figure}

\paragraph{Training algorithm} The signature method of training in deep learning is stochastic gradient descent (SGD) \citep{robbins1951stochastic,rumelhart1986learning}. Trainable parameters, collectively denoted as $\v{\theta}$, are updated in the opposite direction of the gradient of the loss, $\partial L / \partial \v{\theta}$. Intuitively, the $j$-th parameter $\theta_j$ should be reduced by training if the cost function $L$ increases with it; and increased otherwise. For each step of training, since it is usually too expensive to evaluate the loss using the entire training set, the loss is computed using a small number $M$ of randomly selected training examples (a minibatch), indexed by $\mathbb{B}=\{k_1, \cdots,k_M\}$,
\begin{align}
    L_{\rm batch} = \frac{1}{M} \sum_{k\in \mathbb{B}} L(\v{y}^{(k)}, \v{y}_{\rm target}^{(k)}),
\end{align}
hence the name ``stochastic''. For simplicity, we assume a minibatch size of 1 and omit batch in the following equations ($L_{\rm batch}$ will be referred to as $L$, etc.). The gradient, $\partial L / \partial \v{\theta}$ is the direction of parameter change that would lead to maximum increase in the loss function when the change is small enough. To decrease the loss, trainable parameters are updated in the opposite direction of the gradient, with a magnitude proportional to the learning rate $\eta$,
\begin{align}
    \Delta \v{\theta} = - \eta \frac{\partial L}{\partial \v{\theta}}.
    \label{eq:delta-rule}
\end{align}
Parameters such as $\v{W}$ and $\v{b}$ are usually trainable. Other parameters are set by the modelers and called hyperparameters, for example, the learning rate $\eta$. A crucial requirement for computing gradients is differentiability, namely derivatives of functions in the model are well defined.

For a feedforward network without any intermediate (hidden) layer \citep{rosenblatt1962principles} processing a single example $\v{x}$ (minibatch size 1),
\begin{align}
    \v{y} = \v{W}\v{x}+\v{b}, \quad \textrm{or equivalently,} \quad y_i = \sum_j W_{ij}x_j + b_i,
\end{align}
computing the gradient is straightforward,
\begin{align}
    \frac{\partial L}{\partial W_{ij}} = \sum_k\frac{\partial L}{\partial y_k} \frac{\partial y_k}{\partial W_{ij}} = \frac{\partial L}{\partial y_i} x_j,
\end{align}
with $\partial y_k / \partial W_{ij}$ equal to $x_j$ when $k=i$, otherwise 0. In vector notation,
\begin{align}
    \frac{\partial L}{\partial \v{W}} = \frac{\partial L}{\partial \v{y}} \v{x}^\intercal.
\end{align}
Here we follow the convention that $\partial L / \partial \v{W}$ and $\partial L /\partial \v{y}$ have the same form as $\v{W}$ and $\v{y}$, respectively. Assuming that
\begin{align}
    L=\frac{1}{2}\|\v{y} - \v{y}_{\rm target}\|^2 = \frac{1}{2} \sum_j (y_j - y_{{\rm target}, j})^2,
\end{align}
we have,
\begin{align}
    \frac{\partial L}{\partial \v{W}} &= (\v{y} - \v{y}_{\rm target})\v{x}^\intercal, \\
    \Delta W_{ij} & \propto - \frac{\partial L}{\partial W_{ij}} = (y_{{\rm target}, i} - y_i) x_j.
\end{align}
This modification only depends on local information about the input and output units of each connection. Hence, if $y_{{\rm target}, i} > y_i$, $W_{ij}$ should change to increase the net input and $\Delta W_{ij}$ has the same sign as $x_j$. The opposite is true if $y_{{\rm target}, i} < y_i$.

For a multi-layer network, the differentiation is done using the back-propagation algorithm \citep{rumelhart1986learning,lecun88}. To compute the loss $L$, the network is run in a forward pass (Eq. \ref{eq:forwardpass1}-\ref{eq:forwardpass3}). Next, to efficiently compute the exact gradient $\partial L / \partial \v{\theta}$, information about the loss needs to be passed backward, in the opposite direction of the forward pass, hence the name back-propagation.

To illustrate the concept, consider a $N$-layer linear feedforward network (Eq. \ref{eq:forwardpass1}-\ref{eq:forwardpass3}, but with $f(\v{x})=\v{x}$). To compute $\partial L / \partial \v{W}^{(l)}$, we need to compute $\partial L / \partial \v{r}^{(l)}$. From $\v{r}^{(l+1)} = \v{W}^{(l+1)} \v{r}^{(l)} + \v{b}^{(l+1)}$, we have
\begin{align}
\frac{\partial L}{\partial r^{(l)}_i} 
= \sum_j \frac{\partial L}{\partial r^{(l+1)}_j} \frac{\partial r^{(l+1)}_j}{\partial r^{(l)}_i} 
= \sum_j \frac{\partial L}{\partial r^{(l+1)}_j} W^{(l+1)}_{ji}
= \sum_j [W^{(l+1)}]^\intercal_{ij} \frac{\partial L}{\partial r^{(l+1)}_j}.
\end{align}
In vector notation,
\begin{align}
    \frac{\partial L}{\partial \v{r}^{(l)}} 
    = [\v{W}^{(l+1)}]^\intercal\frac{\partial L}{\partial \v{r}^{(l+1)}}
    = [\v{W}^{(l+1)}]^\intercal[\v{W}^{(l+2)}]^\intercal\frac{\partial L}{\partial \v{r}^{(l+2)}}
    = \cdots .
    \label{eq:back-prop}
\end{align} 
Therefore, starting with $\partial L / \partial \v{y}$, $\partial L / \partial \v{r}^{(l)}$ can be recursively computed from $\partial L / \partial \v{r}^{(l+1)}$, for $l = N-1, \cdots, 1$. This computation flows in the opposite direction of the forward pass, and is called the backward pass. In general, back-propagation applies to neural networks with arbitrary differential components.

Computing the exact gradient through back-propagation is considered unrealistic biologically because updating connections at each layer requires precise, non-local information of connection weights at downstream layers (in the form of connection matrix transposed, Eq. \ref{eq:back-prop}).

\subsection{Variations of learning problems/objective functions}
In this and the following sections (\ref{section:variations-architectures}, \ref{section:variations-training}), we introduce common variations of learning problems, network architectures, and training algorithms.

Traditionally, learning problems are divided into three kinds: supervised, reinforcement, and unsupervised learning problems. The difference across these three kinds of learning problems lies in the goal or objective. In supervised learning, each input is associated with a target. The system learns to produce outputs that match the targets. In reinforcement learning, instead of explicit (high-dimensional) targets, the system receives a series of scalar rewards. It learns to produce outputs (actions) that maximize total rewards. Unsupervised learning refers to a diverse set of problems where the system is not provided with explicit targets or rewards. Due to space limitations, we will mainly focus on networks trained with supervised learning in this Primer.

\paragraph{Supervised learning} As mentioned before, for supervised learning tasks, input and target output pairs are provided $\{(\v{x}^{(i)}, \v{y}_{\rm target}^{(i)})\}$. The goal is to minimize the difference between target outputs and actual outputs predicted by the network. In many common supervised learning problems, the target outputs are behavioral outputs. For example, in a typical object classification task, each input is an image containing a single object, while the target output is an integer corresponding to the class of that object (e.g., dog, cat, etc.). In other cases, the target output can directly be neural recording data \citep{mcintosh2016deep,rajan2016recurrent,andalman2019neuronal}.

The classical perceptual decision-making task with random-dot motion \citep{britten1992analysis,roitman02} can be formulated as a supervised learning problem, because there is a correct answer. In this task, animals watch randomly moving dots and report the dots' overall motion direction by choosing one of two alternatives, A or B. This task can be simplified as a network receiving a stream of noisy inputs $x^{(i)}_t$ at every time point $t$ of the $i$-th trial, which can represent the net evidence in support of A and against B. At the end of each trial $t=T$, the system should learn to report the sign of the average input $y_{\rm target}^{(i)} = \mathrm{sign}(\langle x^{(i)}_t\rangle_t)$, $+1$ for choice A and $-1$ for choice B.

\paragraph{Reinforcement learning} For reinforcement learning \citep{sutton2018reinforcement}, a model (an agent) interacts with an environment, such as a (virtual) maze. At time step $t$, the agent receives an observation $\v{o}_t$ from the environment, produces an action $a_t$ that updates the environment state to $\v{s}_{t+1}$, and receives a scalar reward $r_t$ (negative value for punishment). For example, a model navigating a virtual maze can receive pixel-based visual inputs as observations $\v{o}_t$, produce actions $a_t$ that move itself in the maze, and receive rewards when it exits the maze. The objective is to produce appropriate actions $a_t$ given past and present observations that maximize cumulative rewards $\sum_t{r_t}$. In many classical reinforcement learning problems, the observation $\v{o}_t$ equals to the environment state $\v{s}_t$, which contains complete information about the environment.

Reinforcement learning (without neural networks) has been widely used by neuroscientists and cognitive scientists to study value-based learning and decision-making tasks \citep{schultz1997neural,daw2011model,niv2009reinforcement}. For example, in the multi-armed bandit task, the agent chooses between multiple options repeatedly, where each option produces rewards with a certain probability. Reinforcement learning theory can model how the agent's behavior adapts over time, and help neuroscientists study the neural mechanism of value-based behavior.

Deep reinforcement learning trains deep neural networks using reinforcement learning \citep{mnih2015human}, enabling applications to many more complex problems. Deep reinforcement learning can in principle be used to study most tasks performed by lab animals \citep{botvinick2020deep}, since animals are usually motivated to perform the task via rewards. Although many such tasks can also be formulated as supervised learning problems when there exists a correct choice (e.g., perceptual decision making), many other tasks can only be described as reinforcement learning tasks because answers are subjective \citep{haroush2015neuronal,kiani2009representation}. For example, a perceptual decision-making task where there is a correct answer (A, not B) can be extended to assess animals' confidence about their choice \citep{kiani2009representation,song2017reward}. In addition to the two alternatives that result in a large reward for the correct choice and no reward otherwise, monkeys are presented a sure-bet option that guarantees a small reward. Since a small reward is better than no reward, subjects are more likely to choose the sure-bet option when they are less confident about making a perceptual judgement. Reinforcement learning is necessary here because there is no ground-truth choice output: the optimal choice depends on the animals' own confidence level at their perceptual decision.

\paragraph{Unsupervised learning} For unsupervised learning, only inputs $\{\v{x}^{(i)}\}$ are provided, the objective function is defined solely with the inputs and the network parameters $L(\v{x}, \v{\theta})$ (no targets or rewards). For example, finding the first component in Principal Component Analysis (PCA) can be formulated as unsupervised learning in a simple neural network. A single neuron $y$ reading out from a group of input neurons $\v{x}$, $(y=\v{w}^\intercal\v{x})$, can learn to extract the first principle component by maximizing its variance $\mathrm{Var}(y)$ while keeping its connection weights normalized ($\|\v{w}\| = 1$) \citep{oja1982simplified}.

Unsupervised learning is particularly relevant for modeling development of sensory cortices. Although widely-used in machine learning, the kind of labeled data needed for supervised learning, such as image-object class pairs, is rare for most animals. Unsupervised learning has been used to explain neural responses of early visual areas \citep{barlow1961possible,olshausen1996emergence}, and more recently, of higher visual areas \citep{zhuang2019self}.

Compared to reinforcement and unsupervised learning, supervised learning can be particularly effective because the network receives more informative feedback in the form of high-dimensional target outputs. Therefore, it is common to formulate a reinforcement/unsupervised learning problem (or parts of it) as a supervised one. For example, consider an unsupervised learning problem of compressing high-dimensional inputs $\v{x}$ into lower-dimensional representation $\v{z}$ while retaining as much information as possible about the inputs (not necessarily in the information-theoretic sense). One approach to this problem is to train autoencoder networks \citep{rumelhart1986learning, kingma2013auto} using supervised learning. An autoencoder consists of an encoder that maps input $\v{x}$ into a low-dimensional latent representation $\v{z} = f_{\rm encode}(\v{x})$, and a decoder that maps the latent back to a high-dimensional representation $\v{y} = f_{\rm decode}(\v{z})$. To make sure $\v{z}$ contains information about $\v{x}$, autoencoders use the original input as the supervised learning target, $\v{y}_{\rm target} = \v{x}$.

\subsection{Variations of network architectures} \label{section:variations-architectures}
\paragraph{Recurrent neural network}
Besides MLP, another fundamental ANN architecture is recurrent neural networks (RNNs) that process information in time (Figure 2B). In a ``vanilla'' or Elman RNN \citep{elman1990finding}, activity of model neurons at time $t$, $\v{r}_t$, is driven by recurrent connectivity $\v{W}_r$, and by inputs $\v{x}_t$ through connectivity $\v{W}_x$. The output of the network is read out through connections $\v{W}_y$.
\begin{align}
    \v{c}_t & = \v{W}_r \v{r}_{t-1} + \v{W}_x \v{x}_t + \v{b}_r, \\
    \v{r}_t & = f(\v{c}_t), \label{eq:vanillarnn} \\
    \v{y}_t & = \v{W}_y \v{r}_t + \v{b}_y.
\end{align}
Here $\v{c}_t$ represents the cell state, analogous to membrane potential or input current, while $\v{r}_t$ represents the neuronal activity. An RNN can be unrolled in time (Figure 2C) and viewed as a particular form of a MLP,
\begin{align}
    \v{r}_t & = f(\v{W}_r \v{r}_{t-1} + \v{W}_x \v{x}_t + \v{b}_r), \quad {\rm for} \ t=1,\cdots, T.
\end{align}
Here, neurons in the $t$-th layer, $\v{r}_t$ receive inputs from the $(t-1)$-th layer $\v{r}_{t-1}$ and additional inputs from outside of the recurrent network $\v{x}_t$. Unlike regular MLPs, the connections from each layer to the next are shared across time.

Backpropagation also applies to a RNN. While backpropagation in a MLP propagates gradient information from the final layer back (Eq. \ref{eq:back-prop}), computing the gradient for a RNN involves propagating information backward in time (backpropagation-through-time, or BPTT) \citep{werbos1990backpropagation}. Assuming that the loss is computed from outputs at the last time point $T$ and a linear activation function, the key step of backpropagation-through-time is computed similarly to Eq. \ref{eq:back-prop} as
\begin{align}
    \frac{\partial L}{\partial \v{r}_t} 
    = \v{W}_r^\intercal\frac{\partial L}{\partial \v{r}_{t+1}}
    = [\v{W}_r^\intercal]^2 \frac{\partial L}{\partial \v{r}_{t+2}}
    = \cdots .
    \label{eq:bptt}
\end{align}
With an increasing number of time steps in a RNN, weight
modifications involve products of many matrices (Eq. \ref{eq:bptt}). An analogous problem is present for very deep feedforward networks (for example, networks with more than 10 layers). The norm of this matrix product, $\|[\v{W}_r^\intercal]^T\|$, can grow exponentially with $T$, if $\v{W}_r$ is large (more precisely, the largest eigenvalue of $\v{W}_r >  1$); or vanish to zero if $\v{W}_r$ is small, making it historically difficult to train recurrent networks \citep{bengio1994learning,pascanu2013difficulty}. Such exploding and vanishing gradient problems can be substantially alleviated with a combination of modern techniques, including network architectures \citep{hochreiter1997long,he2016deep} and initial network connectivity \citep{le2015simple,he2015delving} that tend to preserve the norm of the backpropagated gradient.

\paragraph{Convolutional neural networks} A particularly important type of network architectures is convolutional neural network (Figure 2D). The use of convolution means that a group of neurons will each process its respective inputs using the same function, in other words, the same set of connection weights. In a typical convolutional neural network processing visual inputs \citep{fukushima1983neocognitron,lecun1990handwritten,krizhevsky2012imagenet,he2016deep}, neurons are organized into $N_{\rm channel}$ ``channels'' or ``feature maps''. Each channel contains $N_{\rm height} \times N_{\rm width}$ neurons with different spatial selectivity. Each neuron in a convolutional layer is indexed by a tuple $i = (i_C, i_H, i_W)$, representing the channel index $(i_C)$, and the spatial preference indices $(i_H, i_W)$. The $i$-th neuron in layer $l$ is typically driven by neurons in the previous layer (bias term and activation function omitted),
\begin{align}
    r_{i_C i_H i_W}^{(l)} = \sum_{j_C j_H j_W} W_{i_C i_H i_W, j_C j_H j_W}^{(l)} r_{j_C j_H j_W}^{(l-1)}.
\end{align}
Importantly, in convolutional networks, the connection weights do not depend on the absolute spatial location of the $i$-th neuron, instead they depend solely on the spatial displacement $(i_H-j_H, i_W-j_W)$ between the pre- and post-synaptic neurons.
\begin{align}
    W_{i_C i_H i_W, j_C j_H j_W}^{(l)} = W_{i_C, j_C}^{(l)}(i_H-j_H, i_W-j_W).
\end{align}
Therefore, all neurons within a single channel process different parts of the input space using the same shared set of connection weights, allowing these neurons to have the same stimulus selectivity with receptive fields at different spatial locations. Moreover, neurons only receive inputs from other neurons with similar spatial preferences, i.e. when $|i_H-j_H|$ and $|i_W-j_W|$ values are small (Figure 2D).

This reusing of weights not only dramatically reduces the number of trainable parameters, but also imposes invariance on processing. For visual processing, convolutional networks typically impose spatial invariance such that objects are processed with the same set of weights regardless of their spatial positions.

In a typical convolutional network, across layers the number of neurons per channel $(N_{\rm height} \times N_{\rm width})$ decreases (with coarser spatial resolution) while more features are extracted (with an increasing number of channels). A classifier is commonly at the end of the system to learn a particular task, such as categorization of visual objects.

\paragraph{Activation function} Most neurons in ANNs, like their biological counterparts, perform nonlinear computations based on their inputs. These neurons are usually point neurons with a single nonlinear activation function $f(\cdot)$ that links the sum of inputs to the output activity. The nonlinearity is essential for the power of ANNs \citep{hornik1989multilayer}. A common choice of activation function is the Rectified Linear Unit (ReLU) function, $f(x)=\max(x,0)$ \citep{glorot2011deep}. The derivative of ReLU at $x=0$ is mathematically undefined, but conventionally set to 0 in practice. ReLU and its variants \citep{clevert2015fast} are routinely used in feedforward networks, while the hyperbolic tangent (tanh) function is often used in recurrent networks \citep{hochreiter1997long}. ReLU and similar activation functions are asymmetric and non-saturating at high value. Although biological neurons eventually saturate at high rate, they often operate in non-saturating regimes. Therefore, traditional neural circuit models with rate units have also frequently used non-saturating activation functions \citep{abbott2005drivers, rubin2015stabilized}.

\paragraph{Normalization} Normalization methods are important components of many ANNs, in particular very deep neural networks \citep{ioffe2015batch,ba2016layer,wu2018group}. Similar to normalization in biological neural circuits \citep{carandini2012normalization}, normalization methods in ANNs keep inputs and/or outputs of neurons in desirable ranges. For example, for inputs $\v{x}$ (e.g., stimulus) to a layer, Layer Normalization \citep{ba2016layer} amounts to a form of ``z-scoring" across units, so that the actual input $\hat{x}_i$ to the $i$-th neuron is 
\begin{align}
    \hat{x}_i & = \gamma \cdot \frac{x_i - \mu}{\sigma} + \beta, \\ \label{eq:normalization_ml}
    \mu & = \langle x_j \rangle, \\
    \sigma & = \sqrt{\langle (x_j - \mu)^2 \rangle + \epsilon}.
\end{align}
\noindent 
where $\langle x_j \rangle$ refers to the average over all units in the same layer; $\mu$ and $\sigma$ are the mean and variance of $\v{x}$.  After normalization, different external inputs lead to the same mean and variance for $\v{\hat{x}}$, set by the trainable parameters $\gamma$ and $\beta$. The values of $\gamma$ and $\beta$ do not depend on the external inputs. The small constant $\epsilon$ ensures that $\sigma$ is not vanishingly small. 

\subsection{Variations of training algorithms} \label{section:variations-training}
\paragraph{Variants of SGD-based methods} Supervised, reinforcement, and unsupervised learning tasks can all be trained with SGD-based methods. Partly due to the stochastic nature of the estimated gradient, directly applying SGD (Eq. \ref{eq:delta-rule}) often leads to poor training performance. Gradually decaying learning rate value $\eta$ during training can often improve performance, since smaller learning rate during late training encourages finer-tuning of parameters \citep{bottou2018optimization}. Various optimization methods based on SGD are used to improve learning \citep{kingma2014adam,sutskever2013importance}. One simple and effective technique is momentum \citep{sutskever2013importance,polyak1964some}, which on step $j$ updates parameters with $\Delta \v{\theta}^{(j)}$ based on temporally smoothed gradients $\v{v}^{(j)}$,
\begin{align}
    \v{v}^{(j)} & = \mu \v{v}^{(j-1)} + \frac{\partial L^{(j)}}{\partial \v{\theta}}, \quad 0 < \mu < 1 \\
    \Delta \v{\theta}^{(j)} & = - \eta \v{v}^{(j)}.
\end{align}
Alternatively, in adaptive learning rate methods \citep{duchi2011adaptive, kingma2014adam}, the learning rate of individual parameter is adjusted based on the statistics (e.g., mean and variance) of its gradient over training steps. For example, in the Adam method \citep{kingma2014adam}, the value of a parameter update is magnified if its gradient has been consistent across steps (low variance). Adaptive learning rate methods can be viewed as approximately taking into account curvature of the loss function \citep{duchi2011adaptive}.

\paragraph{Regularization} Regularization techniques are important during training in order to improve generalization performance by deep networks. Adding a L2 regularization term, $L_{\mathrm{reg}} = \lambda \sum_{ij}W_{ij}^2$, to the loss function \citep{tikhonov1943stability} (equivalent to weight decay \citep{krogh1992simple}) discourages the network from using large connection weights, which can improve generalization by implicitly limiting model complexity. Dropout \citep{srivastava2014dropout} silences a randomly-selected portion of neurons at each step of training. It reduces the network's reliance on particular neurons or a precise combination of neurons. Dropout can be thought of as loosely approximating spiking noise. 

The choice of hyperparameters (learning rate, batch size, network initialization, etc.) is often guided by a combination of theory, empirical evidence, and hardware constraints. For neuroscientific applications, it is important that the scientific conclusions do not rely heavily on the hyperparameter choices. And if they do, the dependency should be clearly documented.

\section{Examples of building ANNs to address neuroscience questions}
In this section, we overview two common usages of ANNs in addressing neuroscience questions.

\subsection{Convolutional networks for visual systems}
Deep convolutional neural networks are currently the standard tools in computer vision research and applications \citep{krizhevsky2012imagenet,simonyan2014very,he2016deep,he2017mask}. These networks routinely consist of tens, sometimes hundreds, of layers of convolutional processing. Effective training of deep feedforward neural networks used to be difficult. This trainability problem has been drastically improved by a combination of innovations in various areas. Modern deep networks would be too large and therefore too slow to run, not to mention train, if not for the rapid development of hardware such as general purpose GPUs (Graphics Processing Units) and TPUs (Tensor Processing Units) \citep{jouppi2017datacenter}. Deep convolutional networks are usually trained with large naturalistic datasets containing millions of high resolution labeled images (e.g., Imagenet \citep{deng2009imagenet}), using training methods with adaptive learning rates \citep{kingma2014adam,tieleman2012lecture}. Besides the default use of convolution, a wide range of network architecture innovations improve performance, including the adoption of ReLU activation function \citep{glorot2011deep}, normalization methods \citep{ioffe2015batch}, and the use of residual connections that can provide an architectural shortcut from a network layer's inputs directly to its outputs \citep{he2016deep}.

\begin{figure}
  \centering
  \includegraphics[width=0.9\linewidth]{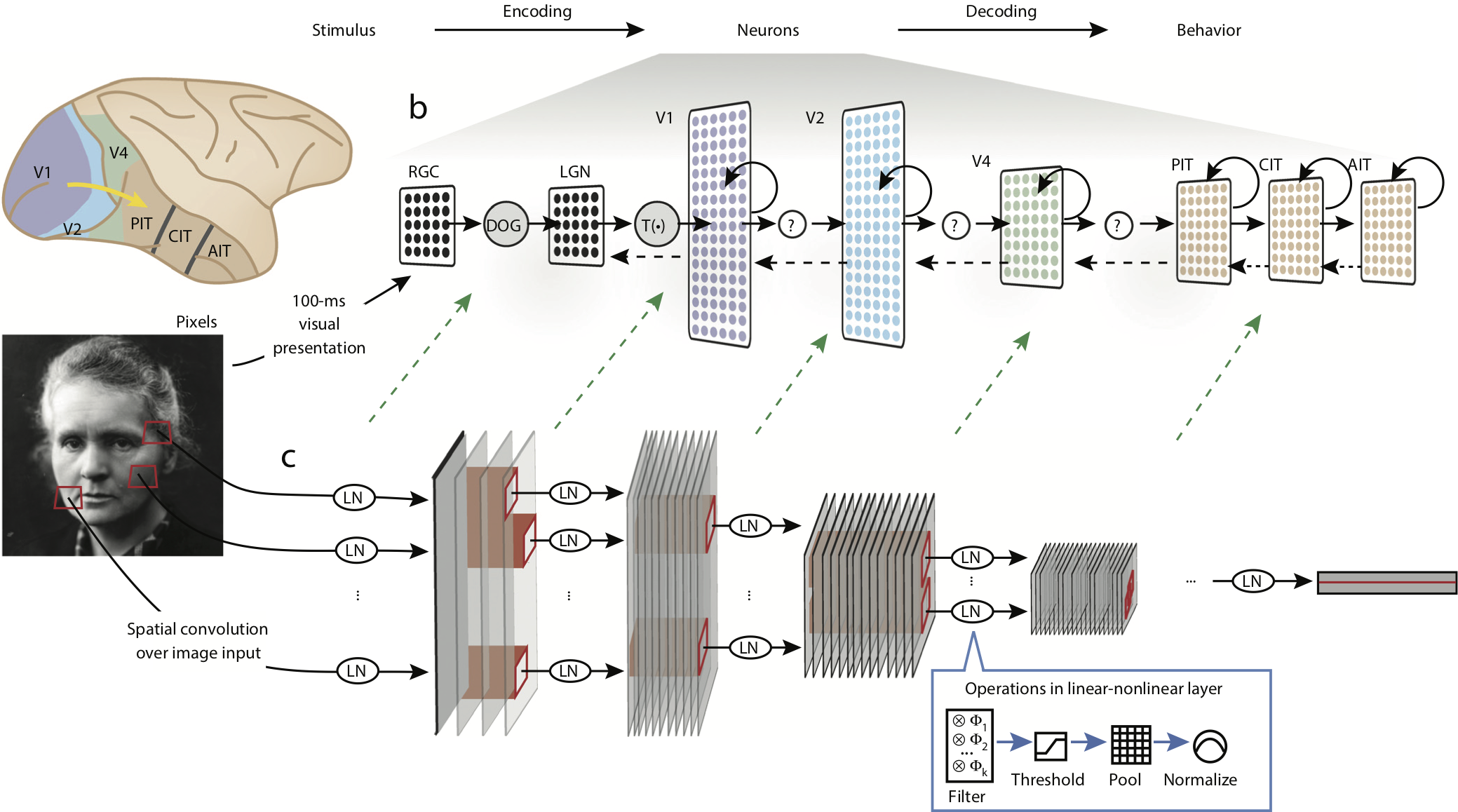}
  \caption{\textbf{Comparing the visual system and deep convolutional neural networks.} The same image is passed through monkey's visual cortex (top) and a deep convolutional neural network (bottom), allowing for side-by-side comparisons between biological and artificial neural networks. Neural responses from IT is best predicted by responses from the final layer of the convolutional network, while neural responses from V4 is better predicted by an intermediate network layer (green dashed arrows). Figure adapted from \cite{yamins2016using}.}
  \label{fig:cnn_schematic}
\end{figure}

Deep convolutional networks have been proposed as computational models of the visual systems, particularly of the ventral visual stream or the ``what pathway'' for visual object information processing (Figure 3) \citep{yamins2016using}. These models are typically trained using supervised learning on the same image classification tasks as the ones used in computer vision research, and in many cases, are the exact same convolutional networks developed in computer vision. In comparison, classical models of the visual systems typically rely on hand-designed features (synaptic weights) \citep{jones1987evaluation,freeman2011metamers,riesenhuber1999hierarchical}, such as Gabor filters, or are trained with unsupervised learning based on the efficient coding principles \citep{barlow1961possible,olshausen1996emergence}. Although classical models have had success at explaining various features of lower-level visual areas, deep convolutional networks surpass them substantially in explaining neural activity in higher-level visual areas in both monkeys \citep{yamins2014performance,cadieu2014deep,yamins2016using} and humans \citep{khaligh2014deep}. Besides being trained to classify objects, convolutional networks can also be trained to directly reproduce patterns of neural activity recorded in various visual areas \citep{mcintosh2016deep,prenger2004nonlinear}.

In a classical work of comparing convolutional networks with higher visual areas \citep{yamins2014performance}, Yamins and colleagues trained thousands of convolutional networks with different architectures on a visual categorization task. To study how similar the artificial and biological visual systems are, they quantified how well the network's responses to naturalistic images can be used to linearly predict responses from the inferior temporal (IT) cortex of monkeys viewing the same images. They found that this neural predictivity is highly correlated with accuracy on the categorization task, suggesting that better IT-predicting models can be built by developing better performing models on challenging natural image classification tasks. They further found that unlike IT, neural responses from the relatively lower visual area, V4, is best predicted by intermediate layers of the networks (Figure 3).

As computational models of visual systems, convolutional networks can model complex, high-dimensional inputs to downstream areas, useful for large-scale models using pixel-based visual inputs \citep{eliasmith2012large}. This process has been made particularly straightforward with the easy access of many pre-trained networks in standard deep learning frameworks like Pytorch \citep{paszke2019pytorch} and Tensorflow \citep{abadi2016tensorflow}.

\subsection{Recurrent neural networks for cognitive and motor systems}
Recurrent neural networks are common machine learning tools to process sequences, such as speech and text. In neuroscience, they have been used to model various aspects of the cognitive, motor, and navigation systems \citep{mante2013context,barak2013fixed,sussillo2015neural,yang2019task,wang2018flexible,cueva2018emergence}. Unlike convolutional networks used to model visual systems that are trained on large-scale image classification tasks, recurrent networks are usually trained on  specific cognitive or motor tasks that neuroscientists are studying. By comparing RNNs trained on the same tasks that animals or humans performed, side-by-side comparisons can be made between RNNs and brains. The comparisons can be made at many levels, including single-neuron activity and selectivity, population decoding, state-space dynamics, and network responses to perturbations. We will expand more on how to analyze RNNs in the next section.

An influential work that uses RNNs to model cognition involves a monkey experiment for context-dependent perceptual decision-making \citep{mante2013context}. In this task, a fraction (called motion coherence) of random moving dots moves in the same direction (left or right); independently a fraction (color coherence) of dots are red, and the rest are green. In a single trial, subjects were cued by a context signal to perform either a motion task (judging the net motion direction is right or left) or a color task (deciding whether there is more red dots than green ones). Monkeys performed the task by temporally integrating evidence for behavioral relevant information (e.g. color) while ignoring the irrelevant feature (motion direction in the color task). Neurons in the prefrontal cortex recorded from behaving animals displayed complex activity patterns, where the irrelevant features are still strongly represented, even though they weakly influence behavioral choices. These counter-intuitive activity patterns were nevertheless captured by a RNN \citep{mante2013context}. Examining the RNN dynamics revealed a novel mechanism by which the irrelevant features are represented, but selectively filtered out and not integrated over time during evidence accumulation. 

To better compare neural dynamics between RNNs and biological systems, RNNs used in neuroscience often treat time differently from their counterparts in machine learning. RNNs in machine learning are nearly always discrete time systems (but see \cite{chen2018neural}), where state at time step $t$ is obtained through a mapping from the state at time step $t-1$ (Eq. \ref{eq:vanillarnn}). The use of a discrete time system means that stimuli that are separated by several seconds in real life can be provided to the network in consecutive time points. To allow for more biologically realistic neural dynamics, RNNs used in neuroscience are often based on continuous time dynamical systems \citep{wilson1972excitatory,sompolinsky1988chaos}, such as
\begin{align}
    \tau \frac{d\v{r}}{dt} = -\v{r}(t) + f(\v{W}_r \v{r}(t) + \v{W}_x \v{x}(t) + \v{b}_r).
\end{align}
Here $\tau$ is the single-unit time scale. This continuous-time system can then be discretized using the Euler method with a time step of $\Delta t (< \tau)$, 
\begin{align}
    \v{r}(t+\Delta t) & \approx \v{r}(t) + \frac{\Delta t}{\tau}[-\v{r}(t) + f(\v{W}_r \v{r}(t) + \v{W}_x \v{x}(t) + \v{b}_r)].
\end{align}

Besides gradient descent through back-propagation, a different line of algorithms has been used to train RNN models in neuroscience \citep{sussillo2009generating,laje2013robust,andalman2019neuronal}. These algorithms are based on the idea of harnessing chaotic systems with weak perturbations \citep{jaeger2004harnessing}. In particular, the FORCE algorithm \citep{sussillo2009generating} allows for rapid learning by modifying the output connections of an RNN to match the target using a recursive least square algorithm. The network output $y(t)$ (assumed to be one-dimensional here) is fed back to the RNN through $\v{w}_{\rm fb}$,
\begin{align}
    \tau \frac{d\v{r}}{dt} & = -\v{r}(t) + f(\v{W}_r \v{r}(t) + \v{W}_x \v{x}(t) + \v{w}_{\rm fb}y(t) + \v{b}_r), \\
    y(t) & = \v{w}^{\intercal}_y \v{r}(t).
\end{align}
Therefore modifying the output connections amounts to a low-rank modification ($\v{w}_{\rm fb}\v{w}^{\intercal}_y$) of the recurrent connection matrix,
\begin{align}
    \tau \frac{d\v{r}}{dt} & = -\v{r}(t) + f([\v{W}_r + \v{w}_{\rm fb}\v{w}^{\intercal}_y] \v{r}(t) + \v{W}_x \v{x}(t)+ \v{b}_r).
\end{align}

\section{Analyzing and understanding ANNs}
Common ANNs used in ML or neuroscience are not easily interpretable. For many neuroscience problems, they may serve better as model systems that await further analyses. Successful training of an ANN on a task does not mean knowing how the system works. Therefore, unlike most ML applications, a trained ANN is not the end goal but merely the prerequisite for analyzing that network to gain understanding.

Most systems neuroscience techniques to investigate biological neural circuits can be directly applied to understand artificial networks. To facilitate side-by-side comparison between artificial and biological neural networks, activity of an ANN can be visualized and analyzed with the same dimensionality reduction tools (e.g., PCA) used for biological recordings \citep{mante2013context,kobak2016demixed,williams2018unsupervised}. To understand causal relationship from neurons to behavior, arbitrary set of neurons can be lesioned \citep{yang2019task}, or inactivated for a short duration akin to optogenetic manipulation in physiological experiments. Similarly, connections between two selected groups of neurons can be lesioned to understand the causal contribution of cross-population interactions \citep{andalman2019neuronal}. 

In this section, we focus on methods that are particularly useful for analyzing ANNs. These methods include optimization-based tuning analysis \citep{erhan2009visualizing}, fixed-point-based dynamical system analysis \citep{sussillo2013opening}, quantitative comparisons between a model and experimental data \citep{yamins2014performance}, and insights from the perspective of biological evolution \citep{lindsey2019unified,richards2019deep}.

\paragraph{Similarity comparison}
Analysis methods such as visualization, lesioning, tuning, fixed-point analysis can offer detailed intuition into neural mechanisms of individual networks. However, with the relative ease of training ANNs, it is possible to train a large amount of neural networks for the same task or dataset \citep{maheswaranathan2019universality,yamins2014performance}. With such volume of data, it is necessary to take advantage of high-throughput quantitative methods that compare different models at scale. Similarity comparison methods compute a scalar similarity score between the neural activity of two networks performing the same task \citep{kriegeskorte2008representational,kornblith2019similarity}. These methods are agnostic about the network form and size, and can be applied to artificial and biological networks alike.

Consider two networks (or two populations of neurons), sized $N_1$ and $N_2$ respectively. Their neural activity in response to the same $D$ task conditions can be summarized by a $D$-by-$N_1$ matrix $\v{R}_1$ and a $D$-by-$N_2$ matrix $\v{R}_2$ (Figure 4A). Representational similarity analysis (RSA) \citep{kriegeskorte2008representational} first computes the dissimilarity or distances of neural responses between different task conditions within each network, yielding a $D$-by-$D$ dissimilarity matrix for each network (Figure 4B). Next, the correlation between dissimilarity matrices of the two networks is computed. A higher correlation corresponds to more similar representations.

Another related line of methods uses linear regression (as used in \citep{yamins2014performance}) to predict $\v{R}_2$ through a linear transformation of $\v{R}_1$, $\v{R_2}\approx \v{W}\v{R_1}$. The similarity corresponds to the correlation between $\v{R}_2$ and its predicted value $\v{W}\v{R_1}$.

\begin{figure}
  \centering
  \includegraphics[width=0.9\linewidth]{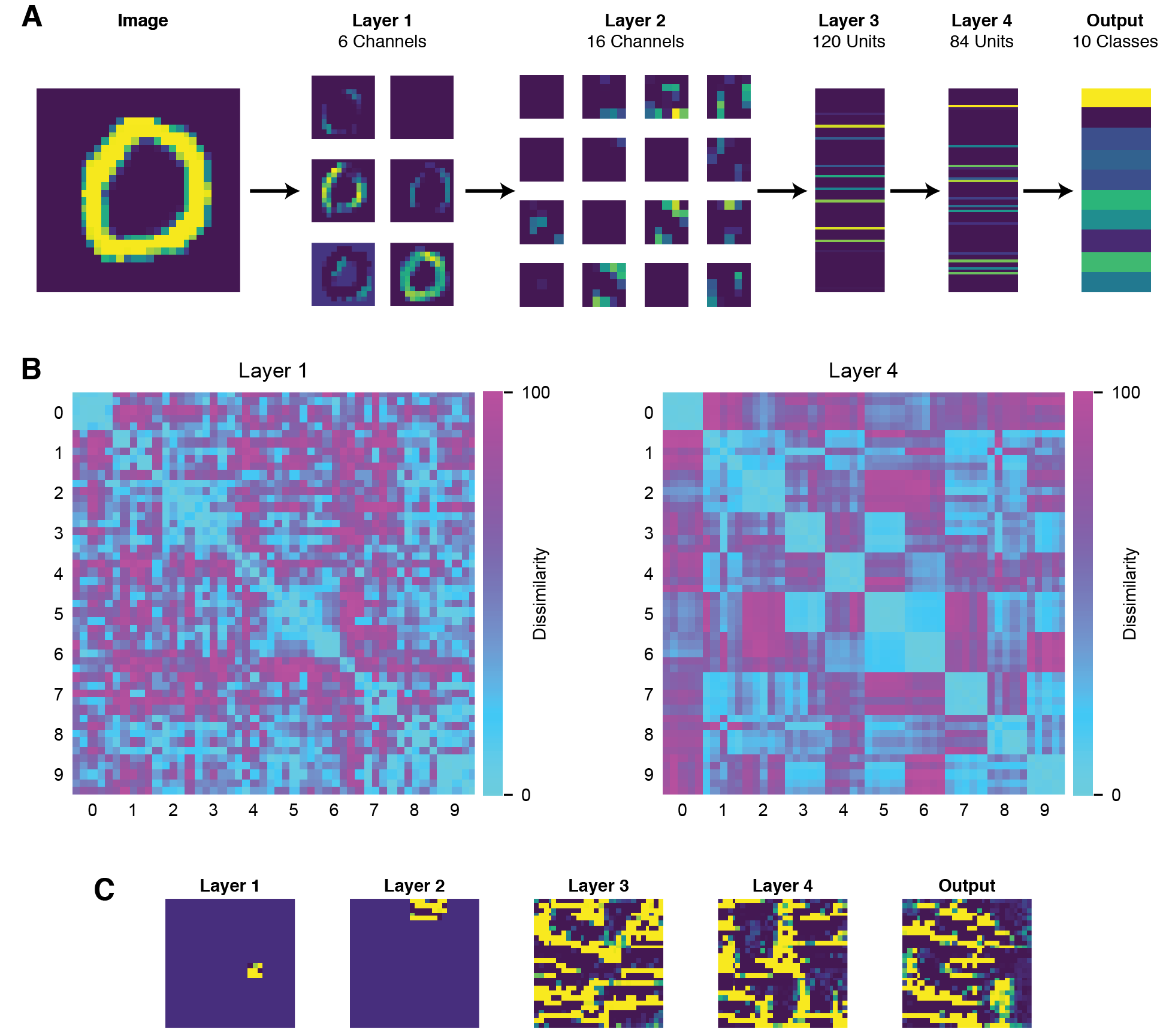}
  \caption{\textbf{Convolutional neural network responses and tuning.} (A) The neural response to an image in a convolutional neural network trained to classify hand-written digits. The network consists of two layers of convolutional processing, followed by two fully-connected layers. (B) Dissimilarity matrices (each $D$-by-$D$) assessing the similar or dissimilar neural responses to different input images. Dissimilarity matrices are computed for neurons in layers 1 and 4 of the network. $D=50$ Images are organized by class (0, 1, etc.), 5 images per class. Neural responses to images in the same class are more similar, i.e. neural representation more category-based, in layer 4 (right) than layer 1 (left). (C) Preferred image stimuli found through gradient-based optimization for sample neurons from each layer. Layers 1 and 2 are convolutional, therefore their neurons have localized preferred stimuli. In contrast, neurons from layers 3 and 4 have non-local preferred stimuli.}
  \label{fig:cnn_response}
\end{figure}

\paragraph{Complex tuning analysis}

Studying tuning properties of single neurons has been one of the most important analysis techniques in neuroscience \citep{kuffler1953discharge}. Classically, tuning properties are studied in sensory areas by showing stimuli parameterized in a low dimensional space (e.g., oriented bars or gratings in vision \citep{hubel1959receptive}). This method is most effective when the neurons studied have relatively simple response properties. A new class of methods treats the mapping of tuning as a high-dimensional optimization problem and directly searches for the stimulus that most strongly activates a neuron. Gradient-free methods such as genetic algorithms have been used to study complex tuning of biological neurons \citep{yamane2008neural}. In deep neural networks, gradient-based methods can be used \citep{erhan2009visualizing,zeiler2014visualizing}. For a neuron with activity $r(\v{x})$ given input $\v{x}$, a gradient-ascent optimization starts with a random $\v{x}_0$, and proceeds by updating the input $\v{x}$ as
\begin{align}
    \v{x} \longrightarrow \v{x} + \Delta \v{x};\hspace{0.2in}  \Delta \v{x} = \eta \frac{\partial r}{\partial \v{x}}.
\end{align}
This method can be used for searching the preferred input to any neuron or any population of neurons in a deep network \citep{erhan2009visualizing,bashivan2019neural}, see Figure 4C for an example. It is particularly useful for studying neurons in higher layers that have more complex tuning properties.

The space of $\v{x}$ may be too high dimensional (e.g., pixel space) for conducting an effective search, especially for gradient-free methods. In that case, we may utilize a lower dimensional space that is still highly expressive. A generative model learns a function that maps a lower-dimensional latent space to a high dimensional space such as pixel space \citep{kingma2013auto,goodfellow2014generative}. Then the search can be conducted instead in the lower-dimensional latent space \citep{ponce2019evolving}.

ANNs can be used to build models for complex behavior that would not be easily done otherwise, opening up new possibilities such as studying encoding of more abstract form of information. For example, \cite{yang2019task} studied neural tuning of task structure, rather than stimuli, in rule-guided problem solving.  An ANN was trained to perform many different cognitive tasks commonly used in animal experiments, including perceptual decision making, working memory, inhibitory control, and categorization. Complex network organization is formed by training, in which recurrent neurons display selectivity for a subset of tasks (Figure 5).

\begin{figure}
  \centering
  \includegraphics[width=0.6\linewidth]{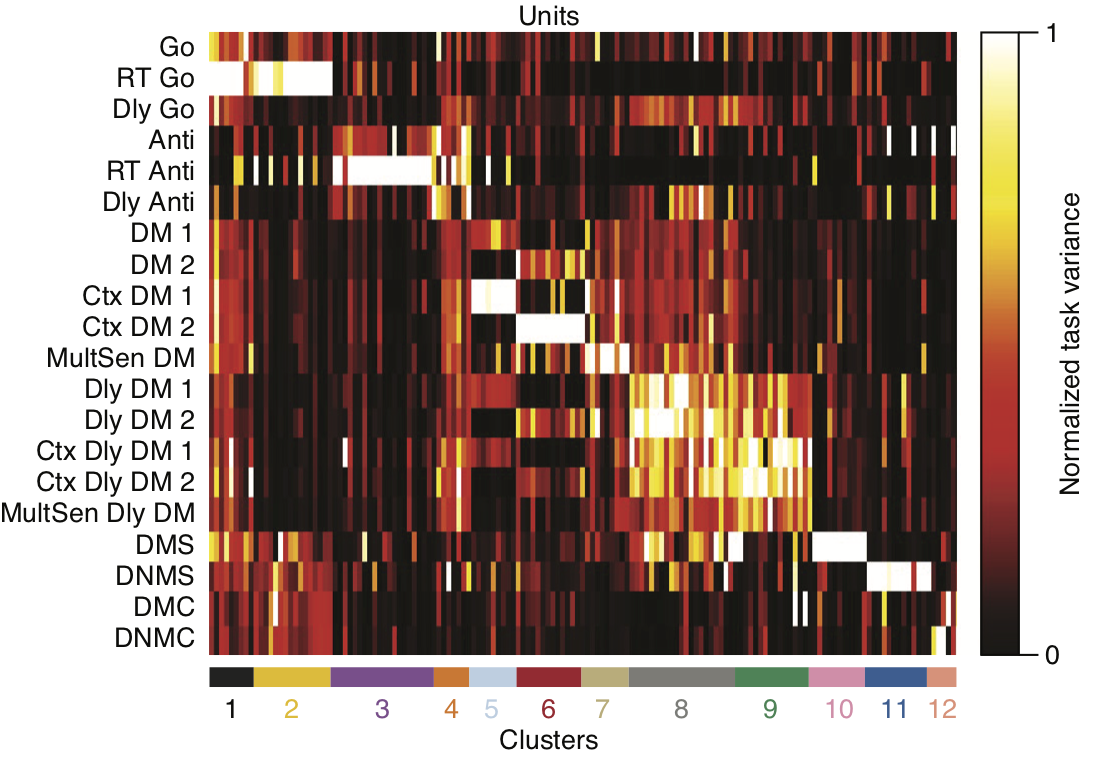}
  \caption{\textbf{Analyzing tuning properties of a neural network trained to perform 20 cognitive tasks.} In a network trained on multiple cognitive tasks, the tuning property of model units to individual task can be quantified. x-axis: recurrent units; y-axis: different tasks. Color measures the degree (between 0 and 1) to which each unit is engaged in a task. Twelve clusters are identified using a hierarchical clustering method (bottom, colored bars). For instance, cluster 3 is highly selective for pro- versus anti-response tasks (Anti) involving inhibitory control; clusters 10 and 11 are involved in delayed match-to-sample (DMS) and delayed non-match-to-sample (DNMS), respectively; cluster 12 is tuned to DMC. Figure adapted from \cite{yang2019task}.}
  \label{fig:tasktuning}
\end{figure}

\paragraph{Dynamical systems analysis}
Tuning properties provide a mostly static view of neural representation and computation. To understand how neural networks compute and process information in time, it is useful to study the dynamics of RNNs \citep{mante2013context,sussillo2013opening,goudar2018encoding,chaisangmongkon2017computing}.

One useful method to understand dynamics is to study fixed points and network dynamics around them \citep{strogatz2001nonlinear}. In a generic dynamical system,
\begin{align}
    \frac{d\v{r}}{dt} = \v{F}(\v{r})
\end{align}
a fixed point $\v{r}_{\mathrm{ss}}$ is a steady state where the state does not change in time, $\v{F}(\v{r}_{\mathrm{ss}})=\v{0}$. The network dynamics at a state $\v{r}=\v{r}_{\mathrm{ss}}+\Delta \v{r}$ around a fixed point $\v{r}_{\mathrm{ss}}$ is approximately linear,
\begin{align}
    \frac{d\v{r}}{dt} = \v{F}(\v{r}) = \v{F}(\v{r}_{\mathrm{ss}}+\Delta \v{r}) \approx \v{F}(\v{r}_{\mathrm{ss}}) + J(\v{r_{\mathrm{ss}}})\Delta \v{r}, \hspace{0.1in}  \frac{d\Delta \v{r}}{dt}=J(\v{r}_\mathrm{ss})\Delta \v{r}.
\end{align}
where $\v{J}$ is the Jacobian of $\v{F}$, $J_{ij}=\partial F_i/\partial r_j$, evaluated at $\v{r}_{\mathrm{ss}}$. This is a linear system which can be understood more easily, for example, by studying the eigenvectors and eigenvalues of $J(\v{r}_\mathrm{ss})$. In ANNs, these fixed points can be found by gradient-based optimization \citep{sussillo2013opening},
\begin{align}
    \mathrm{argmin}_{\v{r}} ||\v{F}(\v{r})||^2.
\end{align}

Fixed points are particularly useful for understanding how networks store memories, accumulate information \citep{mante2013context}, and transition between discrete states \citep{chaisangmongkon2017computing}. 
This point can be illustrated in a network trained to perform a parametric working memory task \citep{romo1999neuronal}. In this task, a sample vibrotactile stimulus at frequency $f_1$ is shown, followed by a delay period of a few seconds; then a test stimulus at frequency $f_2$ is presented, and subjects must decide whether $f_2$ is higher or lower than $f_1$ (Figure 6A). During the delay, neurons in the prefrontal cortex of behaving monkeys showed  persistent activity at a rate that monotonically varies with $f_1$. This parametric working memory encoding emerges from training in an RNN (Figure 6B): in the state-space of this network, neural trajectories during the delay period converge to different fixed points depending on the stored value. These fixed points form an approximate line attractor \citep{seung96} during the delay period (Figure 6C).

\begin{figure}
  \centering
  \includegraphics[width=0.95\linewidth]{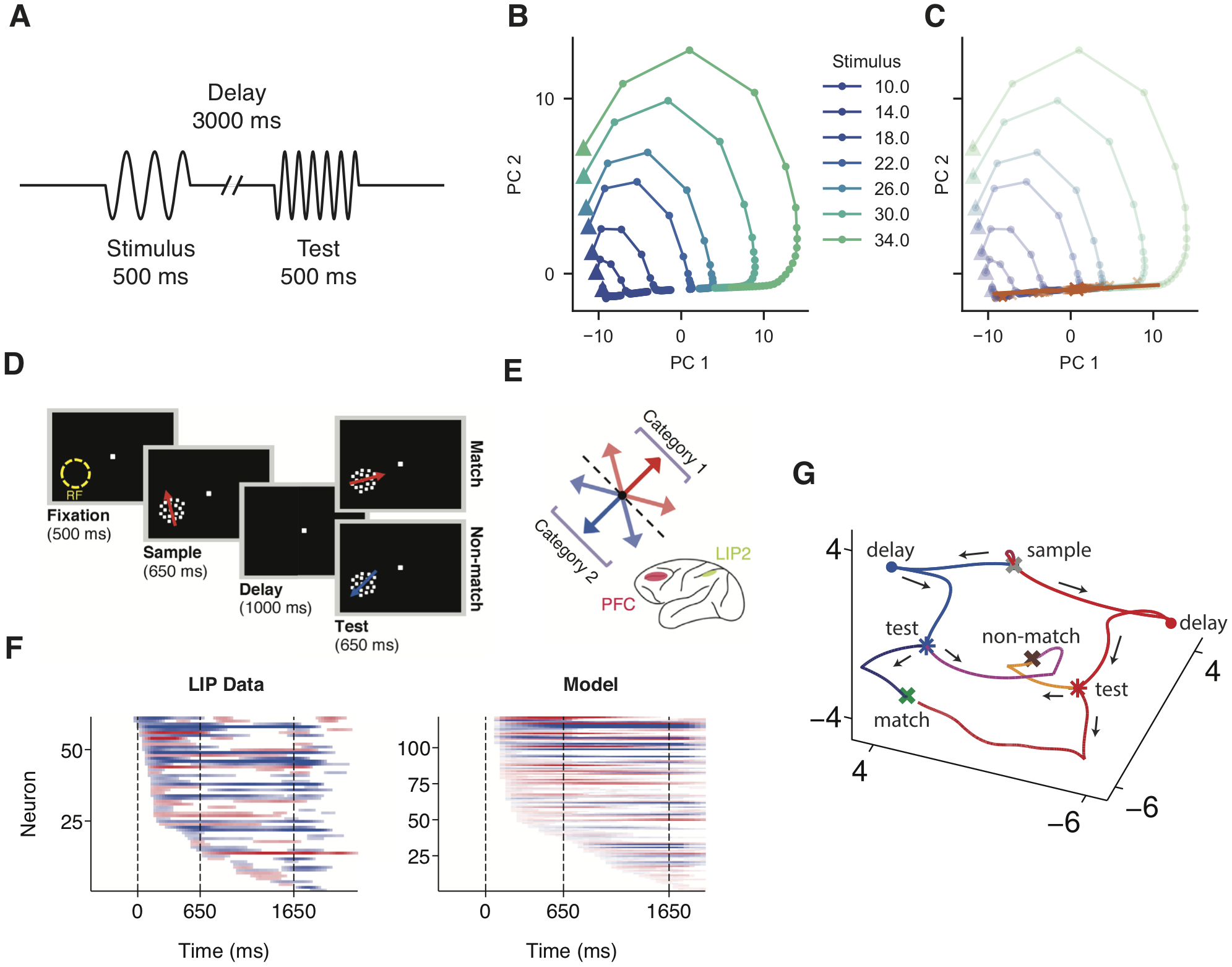}
  \caption{\textbf{Understanding network computation through state-space and dynamical system analysis.} (A-C) In a simple parametric working memory task \citep{romo1999neuronal}, the network needs to memorize the (frequency) value of a stimulus through a delay period (A). The network can achieve such parametric working memory by developing a line attractor (B,C). (B) Trial-averaged neural activity during the delay period in the PCA space for different stimulus values. Triangles indicate the start of the delay period. (C) Fixed points found through optimization (orange cross). The direction of a line attractor can be estimated by finding the eigenvector with a corresponding eigenvalue close to 0. The orange line shows the line attractor estimated around one of the fixed points. (D-G) Training both recurrent neural networks and monkeys on a delayed-match-to-category task \citep{freedman2006experience}. (D) The task is to decide whether the test and sample stimuli (visual moving pattern) belong to the same category. (E) The two categories are defined based on the motion direction of the stimulus (red: category 1; blue: category 2). (F) In a ANN trained to perform this categorization task, the recurrent units of the model display a wide heterogeneity of onset time for category selectivity, similarly to single neurons recorded from monkey posterior parietal cortex (lateral intraparietal area, LIP) during the task. (G) Neural dynamics of a recurrent neural network underlying the performance of the DMC task. The final decision, match (AA or BB) or non-match (AB or BA) corresponds to distinct attractor states located at separate positions in the state space. Similar trajectories of population activity have been found in experimental data. Figure adapted from \cite{chaisangmongkon2017computing}.}
  \label{fig:attractor}
\end{figure}

There is a dearth of examples in computational neuroscience that accounts for not just a single aspect of neural representation or dynamics, but a sequence of computation to achieve a complex task. ANNs offer a new tool to confront this difficulty. \cite{chaisangmongkon2017computing} used this approach to build a model for delayed match-to-category (DMC) tasks. A DMC task (Figure 6D,E) starts with a stimulus sample, say a visual moving pattern, of which a feature (motion direction as an analog quantify from 0 to 360 degrees) is classified into two categories (A in red, B in blue). After a mnemonic delay period, a test stimulus is shown and the task is to decide whether the test has the same category membership as the sample \citep{freedman2006experience}. After training to perform this task, a recurrent neural network shows diverse neural activity patterns similar to parietal neurons in monkeys doing the same task (Figure 6F). The trajectory of recurrent neural population in the state space reveals how computation is carried out through epochs of the task (Figure 6G).

\paragraph{Understanding neural circuits from objectives, architecture, and training}
All above methods seek a mechanistic understanding of ANNs after training. A more integrative view links the three basic ingredients in deep learning: learning problem (tasks/objectives), network architecture, and training algorithm to the solution after training \citep{richards2019deep}. This approach is similar to an evolutionary or developmental perspective in biology, which links environments to functions in biological organisms. It can help explain the computational benefit or necessity of observed structures or functions. For example, compared to purely feedforward networks, recurrently-connected deep networks are better at predicting responses of higher visual area neurons to behaviorally challenging images of cluttered scenes \citep{kar2019evidence}. This suggests a contribution of recurrent connections to classifying difficult images in the brain.

While re-running the biological processes of development and evolution may be difficult, re-training networks with different objectives, architectures, and algorithms is fairly straightforward thanks to recent advances in ML. Whenever training of an ANN leads to a conclusion, it is good practice to vary hyperparameters describing the basic ingredients (to a reasonable degree) to explore the necessary and sufficient conditions for the conclusion \citep{orhan2019diverse,yang2019task,lindsey2019unified}.

The link from the three ingredients to the network solution is typically not rigorous. However, in certain simplified cases, the link can be firmly established by solving the training process analytically \citep{saxe2013exact,saxe2019mathematical}.

\section{Biologically realistic network architectures and learning}
Although neuroscientists and cognitive scientists have had much success with standard neural network architectures (vanilla RNNs) and training algorithms (e.g., SGD) used in machine learning, for many neuroscience questions, it is critical to build network architectures and utilize learning algorithms that are biologically plausible. In this section, we outline methods to build networks with more biologically realistic structures, canonical computations, and plasticity rules.

\subsection{Structured connections}
Modern neurophysiological experiments routinely record from multiple brain areas and/or multiple cell types during the same animal behavior. Computational efforts modeling these findings can be greatly facilitated by incorporating into neural networks fundamental biological structures, such as currently-known cell-type-specific connectivity and long-range connections across model areas/layers.

In common recurrent networks, the default connectivity is all-to-all. In contrast, both local and long-range connectivity in biological neural systems are usually sparse. One way to have a sparse connectivity matrix $\v{W}$ is by element-wise multiplying a trainable matrix $\widetilde{\v{W}}$ with a non-trainable sparse mask $\v{M}$, namely $\v{W} = \widetilde{\v{W}} \odot \v{M}$. To encourage sparsity without strictly imposing it, a L1 regularization term $\beta \sum_{ij}|W_{ij}|$ can be added to the loss function. The scalar coefficient $\beta$ controls the strength of the sparsity constraint.

To model cell-type-specific findings, it is important to build neural networks with multiple cell types. A vanilla recurrent network (Eq. \ref{eq:vanillarnn}) (or any other network) can be easily modified to obey Dale's law by separating excitatory and inhibitory neurons \citep{song2016training},
\begin{align}
    \frac{d\v{r}^E}{dt} & = -\v{r}^E+f_E(\v{W}_{EE} \v{r}^E - \v{W}_{EI} \v{r}^I + \v{W}_{Ex} \v{x} + \v{b}^E), \\
    \frac{d\v{r}^I}{dt} & = -\v{r}^I+f_I(\v{W}_{IE} \v{r}^E - \v{W}_{II} \v{r}^I + \v{W}_{Ix} \v{x} + \v{b}^I),
\end{align}
where an absolute function $|\cdot|$ constrains signs of the connection weights, e.g, $\v{W}_{EE} = |\widetilde{\v{W}}_{EE}|$. After training an ANN to perform the classical ``random dot'' task  of motion direction discrimination \citep{roitman02}, one can ``open the black box'' \citep{sussillo2013opening} and examine the resulting ``wiring diagram'' of recurrent network connectivity pattern (Figure 7). With the incorporation of the Dale's law, the connectivity emerging from training is a heterogeneous version of a biologically-based structured network model of decision-making \citep{wang2002probabilistic}, demonstrating that machine learning brought closer to brain's hardware can indeed be used to shed insights into biological neural networks.

\begin{figure}
  \centering
  \includegraphics[width=0.5\linewidth]{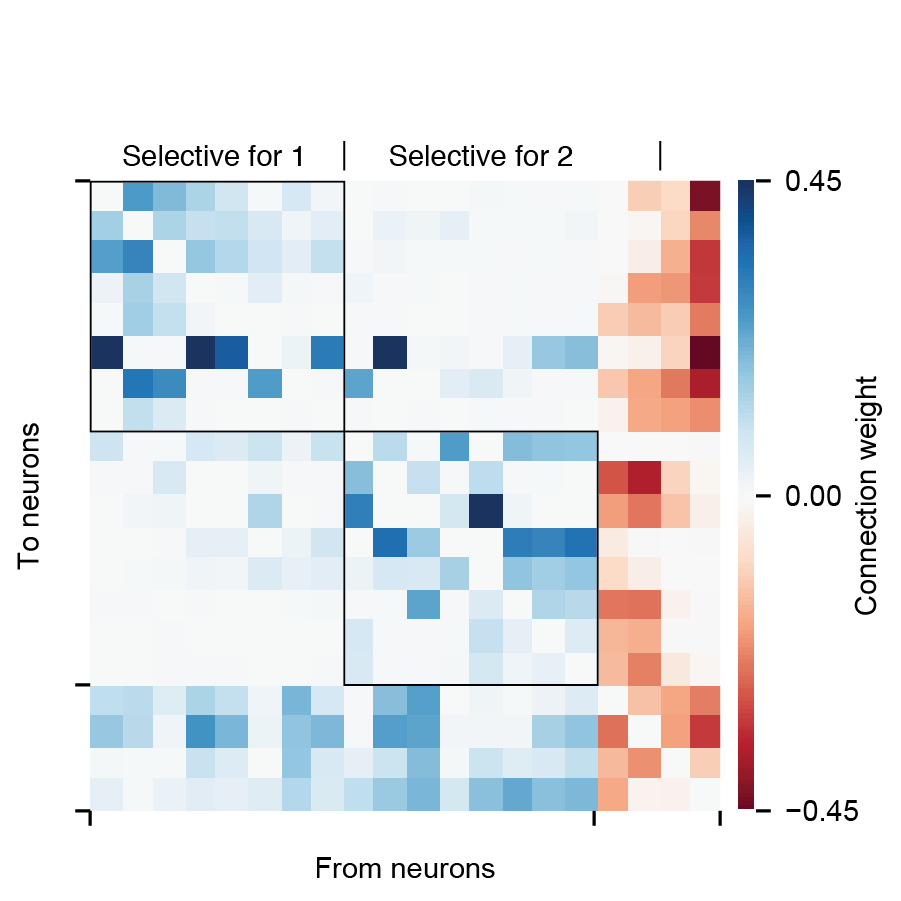}
  \caption{\textbf{Training a network with Dale's law.} Connectivity matrix for a recurrent network trained on a perceptual decision making task. The network respects Dale's law with separate groups of excitatory (blue) and inhibitory (red) neurons. Only connections between neurons with high stimulus selectivity are shown. Neurons are sorted based on their stimulus selectivity to choice 1 and 2. Recurrent excitatory connections between neurons selective to the same choice are indicated by two black squares. Figure inspired from \cite{song2016training}.}
  \label{fig:connectivity}
\end{figure}

The extensive long-range connectivity across brain areas \citep{felleman1991distributed,markov14,oh2014mesoscale} can be included in ANNs. In classical convolutional neural networks \citep{lecun1990handwritten,krizhevsky2012imagenet}, each layer only receives feedforward inputs from the immediate preceding layer. However, in some recent networks, each layer also receives feedforward inputs from much earlier layers \citep{huang2017densely,he2016deep}. In convolutional recurrent networks, neurons in each layer further receive feedback inputs from later layers and local recurrent connections \citep{nayebi2018task,kietzmann2019recurrence}.

\subsection{Canonical computation}
Neuroscientists have identified several canonical computations that are carried out across a wide range of brain areas, including attention, normalization, and gating. Here we discuss how such canonical computations can be introduced into neural networks. They function as modular architectural components that can be plugged into many networks. Interestingly, canonical computations mentioned above all have their parallels in ML-based neural networks. We will highlight the differences and similarities between  purely ML implementations and more biological ones.

\paragraph{Normalization}
Divisive normalization is widely observed in biological neural systems \citep{carandini2012normalization}. In divisive normalization, activation of a neuron $r_i$ is no longer determined by its immediate input $I_i$, $r_i=f(I_i)$. Instead, it is normalized by the sum of inputs $\sum_j I_j$ to a broader pool of neurons called the normalization pool,
\begin{align}
    r_i = f(\gamma\frac{I_i}{\sum_j I_j + \sigma}).
\end{align}
The specific choice of a normalization pool depends on the system studied. Biologically, although synaptic inputs are additive in the drive to neurons, feedback inhibition can effectively produce normalization \citep{ardid07}. This form of divisive normalization is differentiable. So it can be directly incorporated into ANNs.

Normalization is also a critical part of many neural networks in machine learning. Similar to divisive normalization, ML-based normalization methods \citep{ioffe2015batch,ba2016layer,ulyanov2016instance,wu2018group} aim at putting neuronal responses into a range appropriate for downstream areas to process. Unlike divisive normalization, the mean inputs to a pool of neurons is usually subtracted from, instead of dividing, the immediate input (Eq. \ref{eq:normalization_ml}). These methods also compute the standard deviation of inputs to the normalization pool, a step that may not be biologically plausible. Different ML-based normalization methods are distinguished based on their choice of a normalization pool.

\paragraph{Attention}
Attention has been extensively studied in neuroscience \citep{desimone1995neural,carrasco2011visual}. Computational models are able to capture various aspects of bottom-up \citep{koch1987shifts} and top-down attention \citep{reynolds2009normalization}. In computational models, top-down attention usually takes the form of a multiplicative gain field to the activity of a specific group of neurons. In the case of spatial attention, consider a group of neurons, each with a preferred spatial location $x_i$, and pre-attention activity $\widetilde{r}(x_i)$ for a certain stimulus. The attended spatial location $x_q$ results in attentional weights $\alpha_i(x_q)$, which is higher if $x_q$ is similar to $x_i$. The attentional weights can then be used to modulate the neural response of neuron $i$, $r_i(x_q) = \alpha_i(x_q)\widetilde{r}(x_i)$. Similarly, feature attention strengthens the activity of neurons that are selective to the attended features (e.g., specific color). Such top-down spatial and feature attention can be included in convolutional neural networks \citep{lindsay2018biological,yang2018dataset}.

Meanwhile, attention has become widely used in machine learning \citep{bahdanau2014neural,xu2015show,lindsay2020attention}, constituting a standard component in recent natural language processing models \citep{vaswani2017attention}. Although the ML attention mechanisms appear rather different from attention models in neuroscience, as we will show below, the two mechanisms are very closely related.

In deep learning, attention can be viewed as a differentiable dictionary retrieval process. A regular dictionary stores a number of key-value pairs (e.g. word-explanation pairs) $\{(\v{k}^{(i)}, \v{v}^{(i)})\}$, similar to looking up explanation ($\v{v}^{(i)}$) of a word ($\v{k}^{(i)}$). For a given query $\v{q}$, using a dictionary involves searching for the key $\v{k}^{(j)}$ that matches $\v{q}$, $\v{k}^{(j)} = \v{q}$, and retrieving the corresponding value $\v{y} = \v{v}^{(j)}$. This process can be thought of as modulating each value $\v{v}^{(i)}$ based on an attentional weight $\alpha_i$ that measures the similarity between the key $\v{k}^{(i)}$ and the query $\v{q}$. In the simple binary case, 
\begin{align}
    \alpha_i  = 
\begin{cases}
    1,& \text{if } \v{k}^{(i)} = \v{q}\\
    0,              & \text{otherwise}
\end{cases}
\end{align}
\noindent
which modulated the output as 
\begin{align}
    \v{y} & = \sum_i \alpha_i \v{v}^{(i)}.
\end{align}

In the above case of spatial attention, the $i$-th key-value pair is $(x_i, \widetilde{r}(x_i))$, while the query is the attended spatial location $x_q$. Each neuron's response is modulated based on how similar its preferred spatial location (its value) $x_i$ is to the attended location (the query) $x_q$.

The use of ML attention makes the query-key comparison and the value-retrieval process differentiable. A query is compared with every key vector $\v{k}^{(i)}$ to obtain an attentional weight (normalized similarity score) $\alpha_i$,
\begin{align}
    c_i & = \mathrm{score}(\v{q}, \v{k}^{(i)}), \\
    \alpha_1, \cdots, \alpha_N & = \mathrm{normalize}(c_1, \cdots, c_N), \label{eq:attentionnormalize}
\end{align} 
Here the similarity scoring function can be a simple inner product, $\mathrm{score}(\v{q}, \v{k}^{(i)})=\v{q}^\intercal\v{k}^{(i)}$ \citep{bahdanau2014neural}, and the normalization function can be the softmax function,
\begin{align}
    \alpha_i = \frac{e^{c_i}}{\sum_j e^{c_j}}, \quad \textrm{such that} \quad \sum_i \alpha_i = 1.
\end{align}
The use of a normalization function is critical, as it effectively forces the network to focus on a few key vectors (a few attended locations in the case of spatial attention).

\paragraph{Gating}
An important computation for biological neural systems is gating \citep{abbott2006switches,wang18}. Gating refers to the idea of controlling information flow without necessarily distorting its content. Gating in biological systems can be implemented with various mechanisms. Attention modulation multiplies inputs to neurons by a gain factor, providing a graded mechanism of gating at the level of sensory systems \citep{salinas2000gain,olsen2012gain}. Another form of gating may involve several types of inhibitory neurons \citep{wang04,yang16b}. At the behavioral level, gating often appears to be all or none, as exemplified by effects such as inattentional blindness.

In deep learning, multiplicative gating is essential for popular recurrent network architectures such as LSTM (Long Short-Term-Memory) networks (Eq. \ref{eq:lstm}) \citep{hochreiter1997long,gers00} and GRU (Gated Recurrent Units) networks \citep{cho2014learning,chung2014empirical}. Gated networks are generally easier to train and more powerful than vanilla RNNs. Gating variables dynamically control information flow within these networks through multiplicative interactions. In a LSTM network, there are three types of gating variables. Input and output gates, $\v{g}^i_t$ and $\v{g}^o_t$, control the inputs to and outputs of the cell state $\v{c}_t$, while forget gate $\v{g}^f_t$ controls whether cell state $\v{c}_t$ keeps its memory $\v{c}_{t-1}$.
\begin{equation} \label{eq:lstm}
\begin{aligned}
    \v{g}^f_t &= \sigma_g(\v{W}_f \v{x}_t + \v{U}_f \v{r}_{t-1} + \v{b}_f), \\
    \v{g}^i_t &= \sigma_g(\v{W}_i \v{x}_t + \v{U}_i \v{r}_{t-1} + \v{b}_i), \\
    \v{g}^o_t &= \sigma_g(\v{W}_o \v{x}_t + \v{U}_o \v{r}_{t-1} + \v{b}_o), \\
    \v{c}_t &= \v{g}^f_t \odot \v{c}_{t-1} + \v{g}^i_t \odot \sigma_c(\v{W}_c \v{x}_t + \v{U}_c \v{r}_{t-1} +\v{b}_c), \\
    \v{r}_t &= \v{g}^o_t \odot \sigma_r (\v{c}_t).
\end{aligned}
\end{equation}
Here the symbol $\odot$ denotes the element-wise (Hadamard) product of two vectors of the same length ($\v{z}=\v{x}\odot\v{y}$ means $z_i=x_iy_i$). Gating variables are bounded between 0 and 1 by the sigmoid function $\sigma_g$, which can be viewed as a smooth differentiable approximate of a binary step function. A gate is opened or closed when its corresponding gate value is near 1 or 0 respectively. All the weights ($\v{W}$ and $\v{U}$ matrices) are trained. By introducing these gates, a LSTM can in principle keep a memory in its cell state $\v{c}_t$ indefinitely by having the forget gate $\v{g}_t^f=\v{1}$ and input gate $\v{g}_t^i=\v{0}$ (Figure 8). In addition, the network can choose when to read out from the memory by setting its output gate $\v{g}_t^o=\v{0}$ or $\v{1}$. Despite their great utility to machine learning, LSTMs (and GRUs) cannot be easily related to biological neural circuits. Modifications to LSTMs have been suggested so the gating process could be better explained by neurobiology \citep{costa2017cortical}.

\begin{figure}
  \centering
  \includegraphics[width=0.5\linewidth]{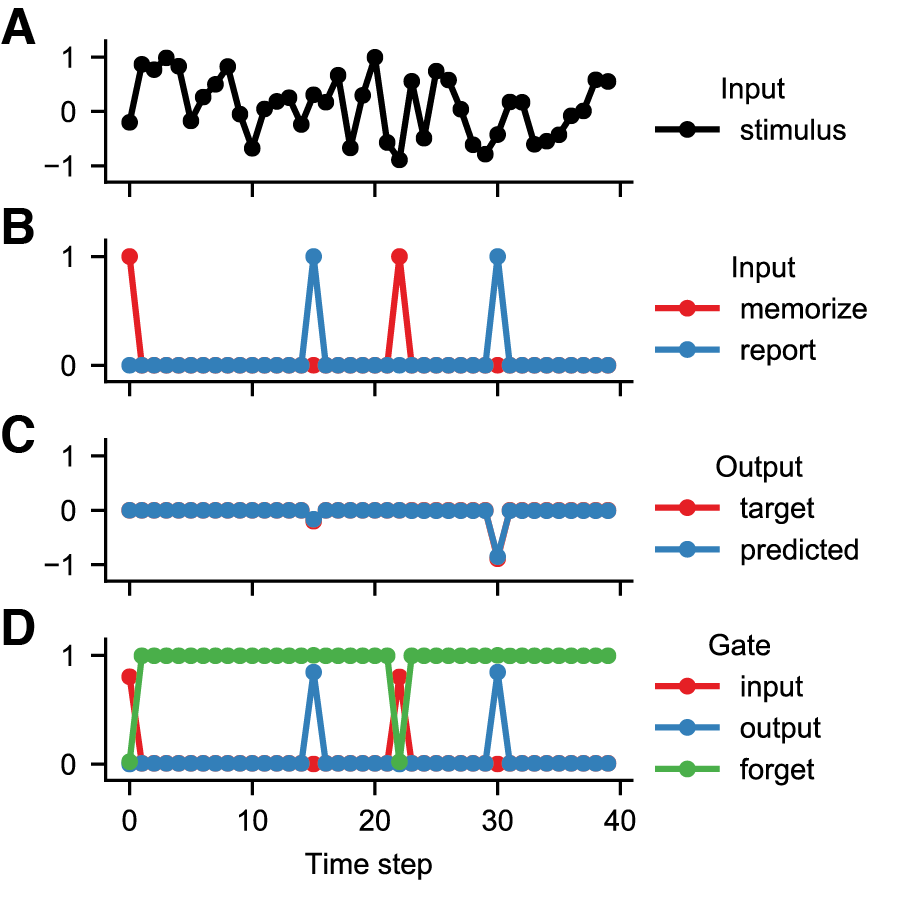}
  \caption{\textbf{Visualizing LSTM activity in a simple memory task.} (A-C) A simple memory task. (A) The network receives a stream of input stimulus, the value of which is randomly and independently sampled at each time point. (B) When the ``memorize input'' (red) is active, the network needs to remember the current value of the stimulus (A), and output that value when the ``report input'' (blue) is next active. (C) After training, a single-unit LSTM can perform the task almost perfectly for modest memory duration. (D) When the memorize input is active, this network opens the input gate (allowing inputs) and closes the forget gate (forgetting previous memory). It opens the output gate when the report input is active.}
  \label{fig:lstmvisualization}
\end{figure}


Although both attention and gating utilize multiplicative interactions, a critical difference is that in attention, the neural modulation is normalized (Eq. \ref{eq:attentionnormalize}), whereas in gating it is not. Therefore, neural attention often has one focus, while neural gating can open or close gates to all neurons uniformly. An important insight from ML is that gating should be plastic, which should inspire neuroscientists to investigate learning to gate in the brain.

\paragraph{Predictive coding}
Another canonical computation proposed for the brain is to compute predictions \citep{rao1999predictive,bastos12,heilbron2018great}. In predictive coding, a neural system constantly tries to make inference about the external world. Brain areas will selectively propagate information that is unpredicted or surprising, while suppressing responses to expected stimuli. To implement predictive coding in ANNs, feedback connections from higher layers can be trained with a separate loss that compares the output of feedback connections with the neural activity in lower layers \citep{lotter2016deep,sacramento2018dendritic}. In this way, feedback connections will learn to predict the activity of lower areas. The feedback inputs will then be used to inhibit neural activity in lower layers.

\subsection{Learning and plasticity}
Biological neural systems are products of evolution, development, and learning. In contrast, traditional ANNs are trained with SGD-based rules mostly from scratch. The back-propagation algorithm of computing gradient descent is well known to be biologically implausible \citep{zipser1988back}. Incorporating more realistic learning processes can help us build better models of brains.

\paragraph{Selective training and continual learning}
In typical ANNs, all connections are trained. However, in biological neural systems, synapses are not equally modifiable. Many synapses can be stable for years \citep{grutzendler2002long,yang2009stably}. To implement selective training of connections, the effective connection matrix $\v{W}$ can be expressed as a sum of a sparse trainable synaptic weight matrix and a non-trainable one, $\v{W} = \v{W}_{\mathrm{train}} + \v{W}_{\mathrm{fix}}$ \citep{rajan2016recurrent,masse2018alleviating}. Or more generally, selective training can be imposed softly by adding to the loss a regularization term $L_{\rm reg}$ that makes it more difficult to change the weights of certain connections,
\begin{align}
    L_{\rm reg} = \beta \sum_{ij} M_{ij} (W_{ij} - W_{\mathrm{fix}, ij})^2.
\end{align}
Here, $M_{ij}$ determine how strongly the connection $W_{ij}$ should stick close to the value $W_{\mathrm{fix}, ij}$.

Selective training of connections through this form of soft constraints has been used by continual learning techniques to combat catastrophic forgetting. The phenomenon of catastrophic forgetting is commonly observed when ANNs are learning new tasks, they tend to rapidly forget previous learned tasks that are not revisited \citep{mccloskey1989catastrophic}. One major class of continual learning methods deals with this issue by selectively training synaptic connections that are deemed unimportant for previously learned tasks or knowledge, while protecting the important ones \citep{kirkpatrick2017overcoming,zenke2017continual}.

\paragraph{Hebbian plasticity}
The predominant idea for biological learning is Hebbian plasticity \citep{hebb2005organization} and its variants \citep{song2000competitive,bi01}. Hebbian plasticity is an unsupervised learning method that drives learning of connection weights without target outputs or rewards. It is essential for classical models of associative memory such as Hopfield networks \citep{hopfield1982neural}, and has a deep link to modern neural network architectures with explicit long-term memory modules \citep{graves2014neural}. 

Supervised learning techniques, especially those based on SGD, can be combined with Hebbian plasticity to develop ANNs that are both more powerful for certain tasks and more biologically realistic. There are two methods to combine Hebbian plasticity with SGD. In the first kind, the effective connection matrix $\v{W} = \widetilde{\v{W}} + \v{A}$ is the sum of two connection matrices, $\widetilde{\v{W}}$ trained by SGD, and $\v{A}$ driven by Hebbian plasticity \citep{ba2016using,miconi2018differentiable},
\begin{align}
    \v{A}(t+1) = \lambda \v{A}(t) + \eta \v{r} \v{r}^\intercal.
\end{align}
Or in component-form,
\begin{align}
    A_{ij}(t+1) = \lambda A_{ij}(t) + \eta r_i r_j.
\end{align}

In addition to training a separate matrix, SGD can be used to learn the plasticity rules itself \citep{bengio1992optimization,metz2018meta}. Here, the plasticity rule is a trainable function of pre- and post-synaptic activity,
\begin{align}
    A_{ij}(t+1) = \lambda A_{ij}(t) + f(r_i, r_j, \v{\theta}).
\end{align}
Since the system is differentiable, parameters $\v{\theta}$, which collectively describe the plasticity rules, can be updated with SGD-based methods. In its simplest form, $f(r_i, r_j, \v{\theta}) = \eta r_i r_j$, where $\v{\theta}=\{\eta\}$. Here, the system can learn to become Hebbian ($\eta > 0$) or anti-Hebbian ($\eta < 0$). Learning of a plasticity rule is a form of meta-learning, using an algorithm (here, SGD) to optimize an inner learning rule (here, Hebbian plasticity).

Such Hebbian plasticity networks can be extended to include more complex synapses with multiple hidden variables in a ``cascade model" of synaptic plasticity \citep{fusi2005cascade}. In theory, properly designed complex synapses can substantially boost a neural network's memory capacity \citep{benna2016computational}. Models of such complex synapses are differentiable, and therefore can be incorporated into ANNs \citep{kaplanis2018continual}.

\paragraph{Short-term plasticity}
In addition to Hebbian plasticity that acts on the time scales from hours to years, biological synapses are subject to short-term plasticity mechanisms operating on the timescale of hundreds of milliseconds to seconds \citep{zucker2002short} that can rapidly modify their effective weights. Classical short-term plasticity rules \citep{mongillo2008synaptic,markram1998differential} are formulated with spiking neurons, but they can be adapted to rate forms. In these rules, each connection weight $w=\widetilde{w}ux$ is a product of an original weight $\widetilde{w}$, a facilitating factor $u$, and a depressing factor $x$. The facilitating and depressing factors are both influenced by the pre-synaptic activity $r(t)$, 
\begin{align}
    \frac{dx}{dt} &= \frac{1-x(t)}{\tau_x} - u(t) x(t) r(t), \\
    \frac{du}{dt} &= \frac{U-u(t)}{\tau_u} + U(1-u(t)) r(t).
\end{align}
High pre-synaptic activity $r(t)$ increases the facilitating factor $u(t)$ and decreases the depressing factor $x(t)$. Again, the equations governing short-term plasticity are fully differentiable, so they can be incorporated into ANNs in the same way as Hebbian plasticity rules \citep{masse2019circuit}.

\cite{masse2019circuit} offers an illustration of how ANNs can be used to test new hypotheses in neuroscience. It was designed to investigate the neural mechanisms of working memory, the brain's ability to maintain and manipulate information internally in the absence of external stimulation. Working memory has been extensively studied in animal experiments using delayed response tasks, in which a stimulus and its corresponding motor response are separated by a temporal gap when the stimulus must be retained internally. Stimulus-selective self-sustained persistent activity during a mnemonic delay is amply documented and considered as the neural substrate of working memory representation \citep{goldman-rakic95,wang01}. However, recent studies suggested that certain short-term memory traces may be realized by hidden variables instead of spiking activity, such as synaptic efficacy that by virtue of short-term plasticity represents past events \citep{stokes2015activity,mongillo2008synaptic}. When an ANN endowed with short-term synaptic plasticity is trained to perform a delayed response task, it does not make an {\it a priori} assumption about whether working memory is represented by hidden synaptic efficacy or neural activity. It was found that activity-silent state can accomplish such a task only when the delay is sufficiently short, whereas persistent activity naturally emerges from training with  delay periods longer than the biophysical time constants of short-term synaptic plasticity. More importantly, training always gives rise to persistent activity, even with a short mnemonic delay period, when information must be manipulated internally, such as mentally rotating a directional stimulus by 90 degrees. This work illustrates how ANNs can contribute to resolving important debates in neuroscience. 

\paragraph{Biologically-realistic gradient descent}
Backpropagation is commonly viewed as biologically unrealistic because the plasticity rule is not local (see Eq. \ref{eq:back-prop}). Efforts have been devoted to approximating gradient descent with algorithms more compatible with the brain's hardware \citep{lillicrap2016random,guerguiev2017towards,roelfsema2018control,lillicrap2020backpropagation}.

In feedforward networks, the backpropagation algorithm can be implemented with synaptic connections feeding back from the final layer \citep{xie2003equivalence}. This implementation assumes that the feedback connections precisely mirror the feedforward connections. This requirement can be relaxed. If a network uses fixed and random feedback connections, the feedforward connections would start to approximately mirror the feedback connections during training (a phenomenon called ``feedback alignment''), allowing for training loss to be decreased \citep{lillicrap2016random}. Another challenge of approximating backpropagation with feedback connections is that the feedback inputs carrying loss information need to be processed differently from feedforward inputs carrying stimulus information. This issue can be addressed by introducing multi-compartmental neurons into ANNs \citep{guerguiev2017towards}. In such networks, feedforward and feedback inputs are processed separately because they are received by the model neurons' soma and dendrites respectively.

These methods of implementing the backpropagation algorithm through synapses propagating information backwards are so far only used for feedforward networks. For recurrent networks, the backpropagation algorithm propagates information backwards in time. Therefore, it is not clear how to interpret the backpropagation in terms of synaptic connections. Instead, approximations can be made such that the network computes approximated gradient information as it runs forward in time \citep{williams1989learning,murray2019local}.

For many neuroscientific applications, it is probably not necessary to justify backpropagation by neurobiology. ANNs often start as ``blank slate", thus training by backpropagation is tasked to accomplish what for the brain amounts to a combination of genetic programming, development and plasticity in adulthood.

\section{Future directions and conclusion}
Recent years have seen a growing impact of ANN models in neuroscience. We have reviewed many of these efforts in the section \textit{Biologically realistic network architectures and learning}. In this final section, we outline other existing challenges and ongoing work to make ANNs better models of brains.

\paragraph{Spiking neural networks}
Most biological neurons communicate with spikes. Harnessing the power of machine learning algorithms for spiking networks remains a daunting challenge. Gradient-descent-based training techniques typically require the system to be differentiable, making it challenging to train spiking networks, because spike generation is non-differentiable. However, several recent methods have been proposed to train spiking networks with gradient-based techniques \citep{courbariaux2016binarized,bellec2018long,zenke2018superspike,nicola2017supervised,huh2018gradient}. These methods generally involve approximating spike generation with a differentiable system during backpropagation \citep{tavanaei2019deep}. Techniques to effectively train spiking networks could prove increasingly important and practical, as neuromorphic hardware that operate naturally with spikes become more powerful \citep{merolla2014million, pei2019towards}.

\paragraph{Standardized protocols for developing brain-like recurrent networks}
In the study of mammalian visual systems, the use of large datasets such as ImageNet \citep{deng2009imagenet} was crucial for producing neural networks that resemble biological neural circuits in the brain. The same has not been shown for most other systems. Although many studies have shown success using neural networks to model cognitive and motor systems, each work usually has its own set of network architectures, training protocols, and other hyperparameters. Simply applying the most common architectures and training algorithms does not consistently lead to brain-like recurrent networks \citep{sussillo2015neural}. Much work remains to be done to search for datasets/tasks, network architectures, and training regimes that can produce brain-resembling artificial networks across a wide range of experimental tasks.

\paragraph{Detailed behavioral and physiological predictions}
Although many studies have reported similarities between brains and ANNs, more detailed comparisons have revealed striking differences \citep{szegedy2013intriguing,henaff2019perceptual,sussillo2015neural}. Deep convolutional networks can achieve similar or better performance on large image classification tasks compared to humans, however, the mistakes they make can be very different from the ones made by humans \citep{szegedy2013intriguing,rajalingham2018large}. It will be important for future ANN models of brains to aim at simultaneously explaining a wider range of physiological and behavioral phenomena.

\paragraph{Interpreting learned networks and learning processes}
With the ease of training neural networks comes the difficulty of analyzing them. Granted, neuroscientists are not foreign to analysis of complex networks, and ANNs are still technologically easier to analyze compared to biological neural networks. However, compared to network models with built-in regularities and small numbers of free parameters, deep neural networks are notoriously complex to analyze and understand, and will likely become even more so as we build more and more sophisticated neural networks. This difficulty is rooted in the use of optimization algorithms to search for parameter values. Since the optimization process in deep learning has no unique optima, the results of optimization necessarily lack the degree of regularities built in hand-designed models. Although we can attempt to understand ANNs from the perspective of its objectives, architectures, and training algorithms \citep{richards2019deep}, which are described with a much smaller number of hyperparameters, the link from these hyperparameters to network representation, mechanism, and behavior is mostly informal and based on intuition.

Despite the difficulties mentioned above, several lines of research hold promise. To facilitate  understanding of learned networks, one can construct variants of neural networks that are more interpretable. For example, low-rank recurrent neural networks utilize recurrent connectivity matrices with low-dimensional structures \citep{mastrogiuseppe2018linking}, allowing for a more straightforward mapping from network connectivity to dynamics and computation.

The dynamics of learning in neural networks can be studied analytically in deep linear networks \citep{saxe2013exact} and very wide nonlinear networks, i.e. networks with a sufficiently large number of neurons per layer \citep{jacot2018neural}. In another line of work, the Information Bottleneck theory proposes that learning processes in neural networks are characterized by two phases, the first extracts information for output tasks (prediction), and the second discards (excessive) information about inputs (compression) \citep{shwartz2017opening}, see also \citep{saxe2019information}. Progress in these directions could shed light on why neural networks can generalize to new data despite having many parameters, which would traditionally indicate over-fitting and poor generalization performance.

\paragraph{Conclusion} Artificial neural networks present a novel approach in computational neuroscience. They have already been used, with certain degree of success, to model various aspects of sensory, cognitive, and motor circuits. Efforts are underway to make ANNs more biologically relevant and applicable to a wider range of neuroscientific questions. In a sense, instead of being viewed as computational models, ANNs can be studied as model systems like fruit flies, mice, and monkeys, but are easily carried out to explore new task paradigms and computational ideas. Of course, one can be skeptical about ANNs as model systems, on the ground that they are not biological organisms. However, computational models span a wide range of biological realism; there should be no doubt that brain research will benefit from enhanced interactions with machine learning and artificial intelligence. In order for ANNs to have a broad impact in neuroscience, it will be important to devote our efforts in two areas. First, we should continue to bring ANNs closer to neurobiology. Second, we should endeavour to ``open the black box'' thoroughly after learning to identify neural representation, temporal dynamics, and network connectivity that emerge from learning, leading to testable insights and predictions by neurobiological experiments. Recurrent neural dynamics emphasized in this Primer represent a salient feature of the brain, further development of strongly recurrent ANNs will contribute to acceleration of progress in neuroscience.

{\bf Acknowledgments}: We thank Vishwa Goudar and Jacob Portes for helpful comments on a draft of this paper. This work was supported by the Simons Foundation, NSF NeuroNex Award DBI-1707398 and the Gatsby Charitable Foundation to GRY; the ONR grant N00014 and Simons Collaboration in the Global Brain (SCGB) (grant 543057SPI) to XJW.

\newpage
\bibliography{references.bib}

\end{document}